\documentclass[prl,nofootinbib,twocolumn,superscriptaddress,preprintnumbers,balancelastpage,longbibliography]{revtex4-1}

\usepackage{graphicx}
\usepackage{color}
\usepackage{hepunits}

\usepackage[english]{babel}

\newcommand{\abs}[1]{\left\lvert #1 \right\rvert}

\newcommand{\es}[2] {\begin{equation} \label{#1} \begin{split} #2 \end{split} \end{equation}}

\newcommand{\Fig}[1]{Fig.~\ref{#1}}

\newcommand{\be}{\begin{equation}}
\newcommand{\ee}{\end{equation}}

\newcommand{\F}{\widetilde{F}}
\newcommand{\E}{\mathbf{E}}
\newcommand{\B}{\mathbf{B}}

\newcommand{\rhoDM}{\rho_{{\rm DM}}}
\newcommand{\g}{g_{a \gamma \gamma}}

\begin{document}
\title{Broadband and Resonant Approaches to Axion Dark Matter Detection}
\author{Yonatan Kahn}
\email{ykahn@princeton.edu}
\affiliation{Department of Physics, Princeton University, Princeton, NJ 08544, U.S.A.}
\author{Benjamin R. Safdi}
\email{bsafdi@mit.edu}
\affiliation{Center for Theoretical Physics, Massachusetts Institute of Technology, Cambridge, MA 02139, U.S.A.}
\author{Jesse Thaler}
\email{jthaler@mit.edu}
\affiliation{Center for Theoretical Physics, Massachusetts Institute of Technology, Cambridge, MA 02139, U.S.A.}

\preprint{MIT-CTP 4763, PUPT 2497}

\date{\today}

\begin{abstract}
When ultralight axion dark matter encounters a static magnetic field, it sources an effective electric current that follows the magnetic field lines and oscillates at the axion Compton frequency.  We propose a new experiment to detect this axion effective current.  In the presence of axion dark matter, a large toroidal magnet will act like an oscillating current ring, whose induced magnetic flux can be measured by an external pickup loop inductively coupled to a SQUID magnetometer.  We consider both resonant and broadband readout circuits and show that a broadband approach has advantages at small axion masses.  We estimate the reach of this design, taking into account the irreducible sources of noise, and demonstrate potential sensitivity to axion-like dark matter with masses in the range of $10^{-14}-10^{-6}~\eV$. In particular, both the broadband and resonant strategies can probe the QCD axion with a GUT-scale decay constant.
\end{abstract}

\maketitle

A broad class of well-motivated dark matter (DM) models consists of light pseudoscalar particles $a$ coupled weakly to electromagnetism \cite{Preskill:1982cy,Abbott:1982af,Dine:1982ah}.  The most famous example is the QCD axion~\cite{Peccei:1977hh,Peccei:1977ur,Weinberg:1977ma,Wilczek:1977pj}, which was originally proposed to solve the strong $CP$ problem. More generally, string compactifications often predict a large number of axion-like particles (ALPs) \cite{Svrcek:2006yi}, with Planck-suppressed couplings to electric (${\bf E}$) and magnetic (${\bf B}$) fields of the form $a \, {\bf E} \cdot {\bf B}$. Unlike QCD axions, generic ALPs do not necessarily couple to the QCD operator $G \tilde G$, where $G$ is the QCD field strength.  The masses and couplings of ALP DM candidates are relatively unconstrained by theory or experiment (see Refs.\ \cite{Essig:2013lka,Marsh:2015xka,Graham:2015ouw} for reviews).  It is therefore important to develop search strategies that cover many orders of magnitude in the axion parameter space.

The ADMX experiment~\cite{Asztalos:2001tf,PhysRevLett.104.041301,Shokair:2014rna} has already placed stringent constraints on axion DM in a narrow mass range around $m_a \sim \text{few} \times 10^{-6}~\eV$.  However, ADMX is only sensitive to axion DM whose Compton wavelength is comparable to the size of the resonant cavity. For the QCD axion, the axion mass $m_a$ is related to the Peccei-Quinn (PQ) symmetry-breaking scale $f_a$ via
\es{axion_mass}{
f_a m_a \simeq f_\pi m_\pi,
}
where $m_\pi \approx 140~\MeV$ ($f_\pi \approx 92~\MeV$) is the pion mass (decay constant).  Lighter QCD axion masses therefore correspond to higher-scale axion decay constants $f_a$. The GUT scale ($f_a \sim 10^{16}~\GeV$, $m_a \sim 10^{-9}~\eV$) is particularly well motivated, but well beyond the reach of ADMX as such small $m_a$ would require much larger cavities. More general ALPs can also have lighter masses and larger couplings than in the QCD case.

In this Letter, we propose a new experimental design for axion DM detection that targets the mass range $m_a \in [10^{-14}, 10^{-6}]$~eV.  Like ADMX, this design exploits the fact that axion DM, in the presence of a static magnetic field, produces response electromagnetic fields that oscillate at the axion Compton frequency.  Whereas ADMX is based on resonant detection of a cavity excitation, our design is based on either broadband or resonant detection of an oscillating magnetic flux with sensitive magnetometers, sourced by an axion effective current.  Our static magnetic field is generated by a superconducting toroid, which has the advantage that the flux readout system can be external to the toroid, in a region of ideally zero static field. Crucially, this setup can probe axions whose Compton wavelength is much larger than the size of the toroid. If this experiment were built, we propose the acronym ABRACADABRA, for ``A Broadband/Resonant Approach to Cosmic Axion Detection with an Amplifying B-field Ring Apparatus.''

For ultralight (sub-eV) axion DM, it is appropriate to treat $a$ as a coherent classical field, since large DM number densities imply macroscopic occupation numbers for each quantum state.  Solving the classical equation of motion with zero DM velocity yields
\es{at}{
a(t) = a_0 \sin(m_a t) = {\sqrt{2 \rho_\text{DM} } \over m_a}  \sin(m_a t) \,,
}
where \mbox{$\rho_\text{DM} \approx 0.3~\GeV$$/$cm$^3$} is the local DM density~\cite{Read:2014qva}.\footnote{The local virial DM velocity $v \sim 10^{-3}$ will give small spatial gradients  $\nabla a \propto v$.}  Through the coupling to the QED field strength $F_{\mu \nu}$,
\es{QED}{
\mathcal{L} \supset -\frac{1}{4}\g a F_{\mu \nu}\F^{\mu \nu},
}
a generic axion will modify Maxwell's equations \cite{Sikivie:1983ip}, and Amp\`ere's circuit law becomes 
\es{ampere}{
\nabla \times \B & =  \frac{\partial \E}{\partial t} - \g \left ( \E \times \nabla a - \B \frac{\partial a}{\partial t} \right) \,,
}
with similar modifications to Gauss's law.
For the QCD axion, $\g = g \alpha_{\rm EM} / (2\pi f_a)$, where $\alpha_{\rm EM}$ is the electromagnetic fine-structure constant and $g$ is an $\mathcal{O}(1)$ number equal to $\sim 0.75$ ($-1.92$) for the DFSZ model~\cite{Dine:1981rt,Zhitnitsky:1980tq} (KSVZ model~\cite{Kim:1979if,Shifman:1979if}).  Thus, in the presence of a static magnetic background $\B_0$, there is an axion-sourced effective current

\es{current}{
 {\bf J_\text{eff}} = g_{a \gamma \gamma} \sqrt{2 \rho_\text{DM} } \cos(m_a t) \B_0.
}
This effective current then sources a real magnetic field, oscillating at frequency $m_a$, that is perpendicular to $\B_0$.

\begin{figure}[t]
\includegraphics[scale=0.5]{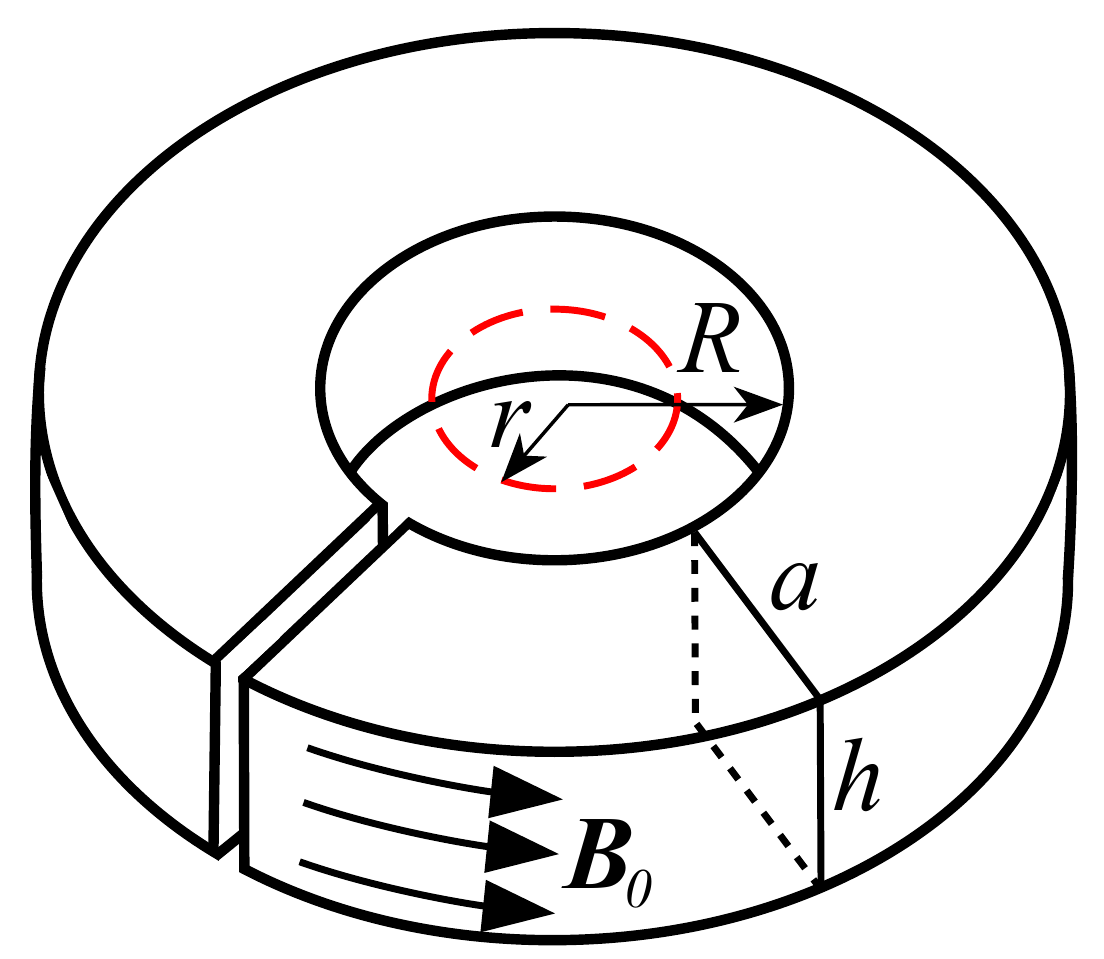}
  \caption{A (gapped) toroidal geometry to generate a static magnetic field $\B_0$. The dashed red circle shows the location of the superconducting pickup loop of radius $r \leq R$. The gap ensures a return path for the Meissner screening current; see discussion in main text.}
\label{fig:toroid}
\vspace{-0.4cm}
\end{figure}

Our proposed design is shown schematically in \Fig{fig:toroid}. The static magnetic field $\B_0$ is generated by a constant current in a superconducting wire wrapping a toroid, and the axion effective current is detected with a superconducting pickup loop in the toroid hole.  In the absence of axion DM (or noise), there is no magnetic flux through the pickup loop.  With axion DM, there will be an oscillating magnetic flux through the pickup loop proportional to $\sqrt{\rho_\text{DM}}$. This design is inspired by cryogenic current comparators (CCCs) \cite{grohmann1974current}, which are used for measuring real currents. The key difference here is the static external field $\B_0$, which generates an effective electric current in the presence of axion DM instead of the real current in the case of the CCC. 

In a real implementation of both designs, the signal flux is actually sourced by a Meissner current which returns along the outside surface of a gapped toroid.  The size of the gap is not crucial for our analysis, but must be sufficiently large that parasitic capacitance effects do not generate a displacement current, which might shunt the Meissner return current and reduce the induced signal $B$-field.  For wires of diameter 1 mm and a meter-sized toroid, a gap of a few millimeters allows unscreened currents up to the frequency at which the magnetoquasistatic approximation breaks down and displacement currents are unavoidable. In what follows, we will estimate our sensitivity using the axion effective current which is correct up to $\mathcal{O}(1)$ geometric factors.

We consider two distinct circuits for reading out the signal, both based on a superconducting quantum interference device (SQUID).  The broadband circuit uses a untuned magnetometer in an ideally zero-resistance setup, while the resonant circuit uses a tuned magnetometer with irreducible resistance.  Both readout circuits can probe multiple orders of magnitude in the axion DM parameter space, though the broadband approach has increased sensitivity at low axion masses.

A related proposal, utilizing the axion effective current, was put forth recently by Ref.~\cite{Sikivie:2013laa} (see also Ref.~\cite{ThomasTalk} for a preliminary proposal and Ref.~\cite{Chaudhuri:2014dla} for a similar design for detecting dark photon DM).  That design was based on a solenoidal magnetic field, with the pickup loop located inside of the solenoid, and focused on resonant readout using an $LC$ circuit.  The design presented here offers a few advantages. First, the toroidal geometry significantly reduces fringe fields compared to a solenoidal geometry.  Second, the pickup loop is located in an ideally zero-field region, outside of the toroidal magnetic field $\B_0$, which should help reduce flux noise. Third, as we will show, broadband readout has significant advantages over resonant readout at low axion masses.  Our proposal is complementary to the recently proposed CASPEr experiment~\cite{Budker:2013hfa}, which probes a similar range of axion masses but measures the coupling to nuclear electric dipole moments rather than the coupling to QED. See Refs.~\cite{Baker:2011na,Graham:2011qk,Graham:2013gfa,Horns:2012jf,Horns:2013ira,Stadnik:2013raa,Beck:2013jha,Beck:2014aqa,Hong:2014vua,Roberts:2014dda,Roberts:2014cga,Stadnik:2014tta,Hill:2015kva,Hill:2015vma,McAllister:2015zcz} for other proposals to detect axion DM.

\begin{figure}[t]
\vspace{-0.2cm}
\includegraphics[width=\columnwidth]{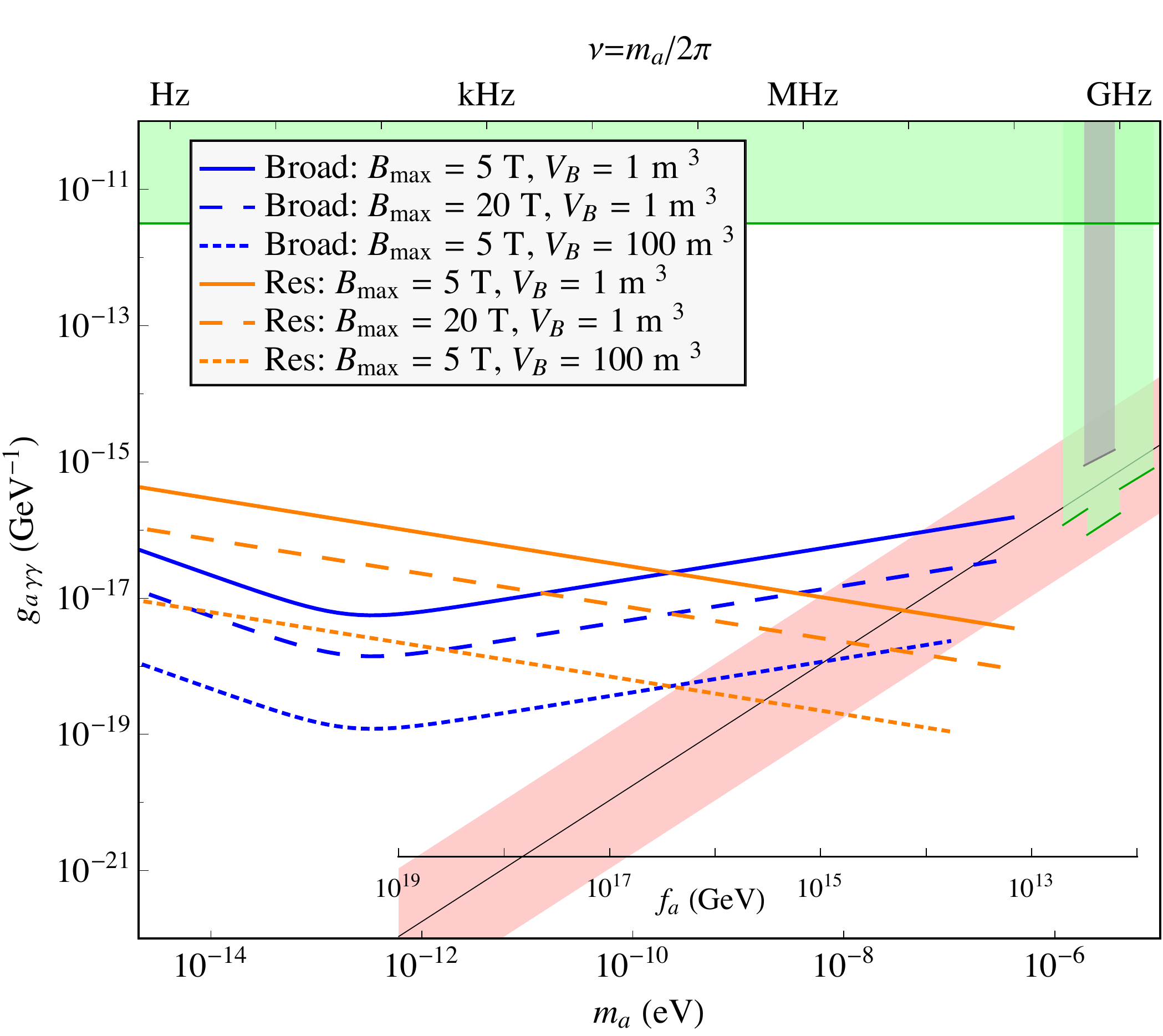}
  \caption{Anticipated reach in the $\g$ vs.\ $m_a$ plane for the broadband (Broad) and resonant (Res) strategies. The benchmark parameters are $T = 0.1~{\rm K}$, $r  = a = R = h/3$ (see \Fig{fig:toroid}), and $L_p = L_{\rm min} \approx \pi R^2/h$. The total measurement time for both strategies is $t = 1$ yr, where the resonant experiment scans from 1 Hz to 100 MHz.  The expected parameters for the QCD axion are shown in shaded red, with the corresponding decay constant $f_a$ inset at bottom right.  The projected sensitivities of IAXO \cite{Vogel:2013bta} and ADMX \cite{Shokair:2014rna} are shown shaded in light green. Published limits from ADMX \cite{PhysRevLett.104.041301} are shown in gray.}
\label{fig:reach}
\vspace{-0.5cm}
\end{figure}

For concreteness, our sensitivity studies are based on a toroid of rectangular cross section (height $h$, width $a$) and inner radius $R$, as illustrated in \Fig{fig:toroid}.  The magnetic field inside the toroid volume is
\es{bfieldconf}{
\B_0(s) = B_{\rm max} \frac{R}{s} \hat{\bm{\phi}},
}
where $s$ is the distance from the central axis of the toroid, $\hat{\bm{\phi}}$ is the azimuthal direction, and $B_{\rm max}$ is the magnitude of $\B_0$ at the inner radius.  The flux through the pickup loop of radius $r \leq R$ can be written as
\es{flux_eqn}{
\Phi_{\rm pickup}(t) = \g \,  B_{\rm max} \, \sqrt{2 \rhoDM} \, \cos(m_a t) \, V_B. 
}
The effective volume containing the external $B$-field is
\es{flux_eqn_V}{
V_B = \int_0^{r} \! \mathrm{d} r' \int_R^{R+a} \! \!\!\!  \mathrm{d} s \int_0^{2\pi} \! \! \mathrm{d} \theta \, \frac{R h r'(s - r' \cos \theta)}{\tilde{r}^2 \sqrt{h^2 + 4\tilde{r}^2}}, }
with \mbox{$\tilde{r}^2 \equiv s^2 + r'^2 - 2 s r' \cos \theta$}. We work in the magnetoquasistatic limit, $2\pi/ m_a \gg r, R, h, a$; at higher frequencies, displacement currents can potentially screen our signal. As an illustration, we consider a meter-sized experiment, where $V_B = 1 \ {\rm m}^3$ for $r = R = a = h/3 = 0.85$ m, with sensitivity to $m_a \lesssim 10^{-6} \ \eV$. For an example of the magnitude of the generated fields, the average $B$-field sourced by a GUT-scale KSVZ axion ($f_a = 10^{16} \ \GeV$) with $V_B = 100 \ {\rm m}^3$ and $B_{\rm max} = 5 \ {\rm T}$ is $2.5 \times 10^{-23} \ {\rm T}$. To detect such a small $B$-field at this frequency, we need a flux noise sensitivity of $1.2 \times 10^{-19} \ {\rm Wb}/\sqrt{\rm Hz}$ for a  measurement time of 1 year in a broadband strategy (see below). The anticipated reach for various $V_B$ and $B_{\rm max}$ is summarized in \Fig{fig:reach}.

\begin{figure}[t]
\includegraphics[scale=0.28]{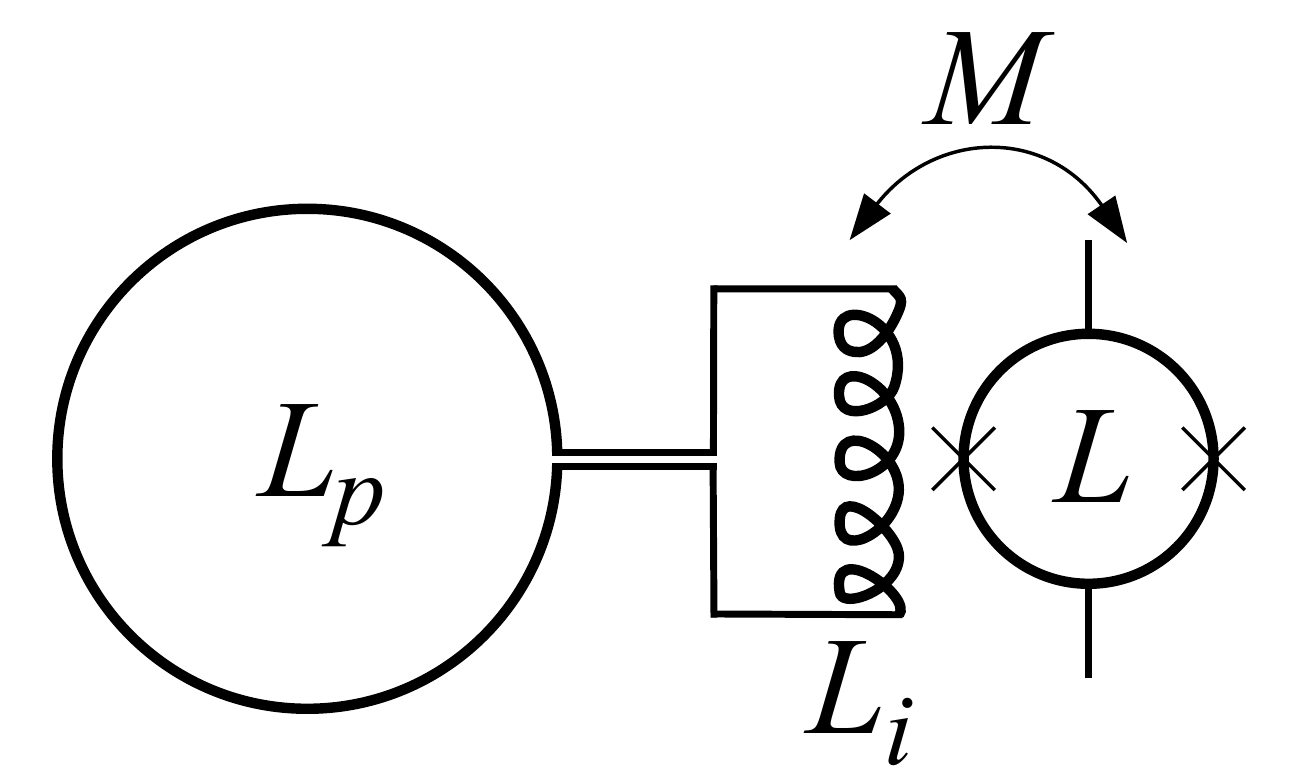}
\includegraphics[scale=0.28]{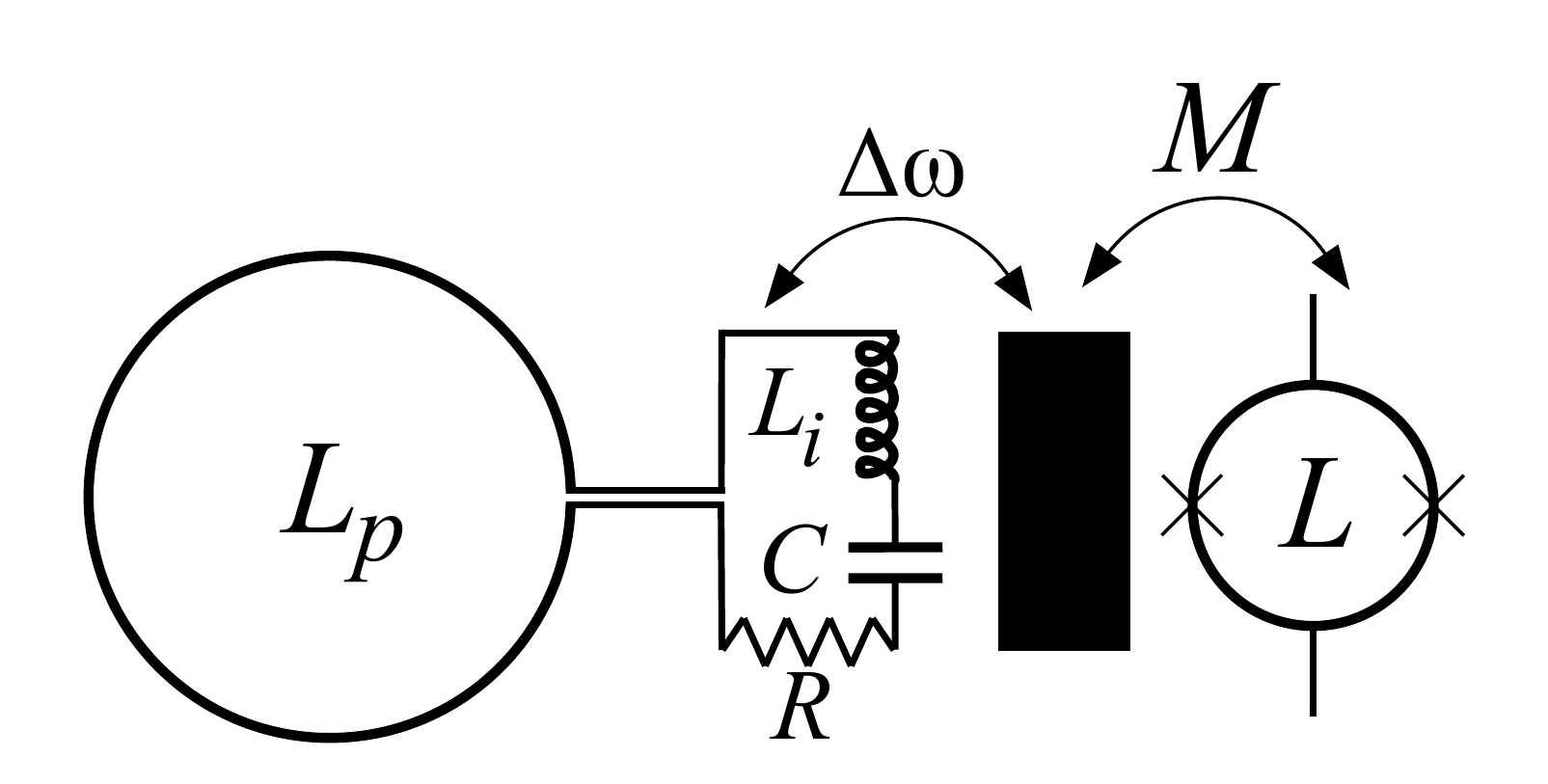}
  \caption{Schematics of our readout circuits. {\bf Left:} broadband (untuned magnetometer). The pickup loop $L_p$ is placed in the toroid hole as in \Fig{fig:toroid} and connected in series with an input coil $L_i$, which has mutual inductance $M$ with the SQUID of self-inductance $L$. {\bf Right:} resonant (tuned magnetometer). $L_p$ is now in series with both $L_i$ and a tunable capacitor $C$.  A ``black box'' feedback circuit modulates the bandwidth $\Delta \omega$ and has mutual inductance $M$ with the SQUID.}
\label{fig:circuits}
\vspace{-0.2cm}
\end{figure}

{\it Broadband approach}---In an untuned magnetometer, a change in flux through the superconducting pickup loop induces a supercurrent in the loop.  As shown in \Fig{fig:circuits} (left), the pickup loop (inductance $L_p$) is connected in series with an input coil $L_i$, which is inductively coupled to the SQUID (inductance $L$) with mutual inductance $M$. The flux through the SQUID is proportional to the flux through the pickup loop and is maximized when $L_i \approx L_p$~\cite{myers2007calculated}:
\be
\label{eq:SQUIDinputflux}
\Phi_{{\rm SQUID}} \approx \frac{\alpha}{2} \sqrt{\frac{L}{L_p}} \Phi_{{\rm pickup}}.
\ee
Here $\alpha$ is an $\mathcal{O}(1)$ number, with $\alpha^2 \approx 0.5$ in typical SQUID geometries \cite{Clarke:1979aa}.

Clearly, the flux through the SQUID will be maximized for $L$ as large as possible and $L_p$ as small as possible.  A typical SQUID has inductance $L = 1 \ {\rm nH}$. A superconducting pickup loop of wire radius $\phi = 1$ mm and loop radius $r = 0.85$ m has geometric inductance of \cite{shoenberg1954superconductivity}
\be
\label{eq:PickupLoopL}
L_p = r (\ln(8r/\phi) - 2) \approx 7 \ \mu{\rm H},
\ee
but this may be reduced with smaller loops in parallel as in a fractional-turn magnetometer \cite{zimmerman1971sensitivity,drung1990low}. The minimum inductance is limited by the magnetic field energy $\frac{1}{2} \int \B^2 \, \text{d}V$ stored in the axion-sourced response field, and is approximately
\be
\label{eq:Lmin}
 L_{\rm min} \approx \pi R^2/h.
 \ee
With a ``tall'' toroid where $h = 3R$, one can achieve $L_{\rm min} \approx 1 \ \mu {\rm H}$ and $\Phi_{{\rm SQUID}} \approx 0.01 \Phi_{{\rm pickup}}$ for $R = 0.85$ m. Since the pickup loop area is much larger than the magnetometer area, the $B$-field felt by the SQUID is significantly enhanced compared to the axion-induced field in the pickup loop. The $B$-field enhancement takes advantage of the fact that we are working in the near-field limit, so that the induced $B$-field adds coherently over the pickup loop.

To assess the sensitivity of the untuned magnetometer to the axion-sourced oscillating flux in~\eqref{flux_eqn}, we must characterize the noise of the circuit. In a pure superconducting circuit at low frequencies, there is zero noise in the pickup loop and input coil, and the only source of noise is in the SQUID, with contributions from thermal fluctuations of both voltage and current. Despite their thermal origin, we will refer to these as ``magnetometer noise'' to distinguish them from noise in the pickup loop circuit (which dominates in the resonant case below).  At cryogenic temperatures ($T \lesssim 60 \ {\rm mK}$), thermal current and voltage noise are subdominant to the current shot noise $S_{J, 0}$ in the SQUID tunnel junctions~\cite{Clarke:1979aa}, which sets an absolute (temperature-independent) floor for the magnetometer noise. See the appendix for a more detailed discussion of noise in a real implementation of this design.

A typical, temperature-independent flux noise for commercial SQUIDs at frequencies greater than $\sim$10 Hz is 
\be
\label{eq:ZeroTFluxNoise}
S_{\Phi,0}^{1/2} \sim 10^{-6} \Phi_0/\sqrt{{\rm Hz}},
\ee
where $\Phi_0 = h / (2e) = 2.1 \times 10^{-15} \ {\rm Wb}$ is the flux quantum.  We use this noise level and a fiducial temperature of 0.1 K as our benchmark. DC SQUIDS are also known to exhibit $1/f$ noise which dominates below about 50 Hz at 0.1 K \cite{anton2013magnetic}. We estimate the reach of our broadband strategy down to 1 Hz assuming $1/f$ noise is the sole irreducible source of noise at these low frequencies, but in a realistic experiment, environmental noise would likely contribute as well; see the Supplementary Material for more details.

Following \cite{Budker:2013hfa}, the signal-to-noise ratio $S/N$ improves with integration time $t$ as
\es{eq:SoverN}{
S / N \sim \abs{\Phi_\text{SQUID}} \, (t \tau)^{1/4} / S_{\Phi,0}^{1/2}
}
for \mbox{$t > \tau$}, where $\tau$ is the axion coherence time.\footnote{When \mbox{$t < \tau$}, \mbox{$S / N \sim |\Phi_\text{SQUID}| \sqrt{t}  / S_{\Phi,0}^{1/2} $}.} The axion coherence time is approximately 
\es{eq:CohTime}{
\tau \sim \frac{2\pi}{m_a v^2} \sim 10^6 \frac{2\pi}{m_a} \approx 3 \times 10^{4}~\text{s} \left({ 10^{-12}~\text{eV} \over m_a} \right)  \,,
}
where we have taken $v \sim 10^{-3}$ as the local DM virial velocity.  We assume a fiducial integration time of $t = 1$ year, so that $t \gg \tau$ over most of the mass range of interest. We also assume a geometry with $r = R = a = h/3$ and a pickup loop inductance $L_p = L_{\rm min}$. Then, requiring $S/N > 1$ after time $t$ implies sensitivity to 
\es{eq:BroadbandReach}{
\g > \ & 6.3 \times 10^{-18} \ \GeV^{-1}  \left(\frac{m_a}{10^{-12} \ \eV} \frac{1 \ {\rm year}}{t} \right)^{1/4} \frac{5 \ {\rm T}}{B_{\rm max}}   \\
& \times  \left(\frac{0.85 \ {\rm m}}{R}\right)^{5/2}  \sqrt{\frac{0.3 \ \GeV/{\rm cm}^3}{\rhoDM}}  \frac{S_{\Phi,0}^{1/2}} {10^{-6} \Phi_0 /\sqrt{\text{Hz}}}.
}

As shown in \Fig{fig:reach}, an ideal broadband setup with the benchmark parameters in Eq.~\eqref{eq:BroadbandReach} could begin to probe the QCD axion band for $f_a \lesssim 10^{14} \ \GeV$, which is not far below the GUT scale. The sensitivity improves for larger magnetic fields or larger toroids; for a toroid with $V_B = 100 \ {\rm m}^3$, one can probe the QCD axion band at the GUT scale.  However, larger experiments may not be sensitive to axion masses near $10^{-6} \ \eV$ because displacement currents may partially cancel the axion-sourced flux.  Note that the sensitivity to $\g$ increases at \emph{smaller} $m_a$, due to the increase in axion coherence time.  

{\it Resonant approach}---We now turn to an analysis of a tuned magnetometer, shown in \Fig{fig:circuits} (right). This readout circuit has the advantage of enhancing the signal by the quality factor $Q$ at the resonant frequency. The tuned circuit is a standard design for detecting small magnetic fields at a given frequency (see e.g.~Ref.~\cite{Clarke:1979aa}).  Similar tuned circuits have been considered before for axion DM detection~\cite{Sikivie:2013laa} and dark-photon DM detection~\cite{Chaudhuri:2014dla}; our analysis follows closely those of Refs.~\cite{Chaudhuri:2014dla} and \cite{myers2007calculated}. 

In a practical implementation of an $LC$ circuit with resonant frequency $\omega = 1/\sqrt{LC}$, the capacitor has nonzero intrinsic resistance $R$.  Therefore, the circuit has a finite bandwidth $\Delta \omega_\text{LC} = \omega/Q_0$, where $Q_0 = (\omega C R)^{-1}$.   To maximize the axion signal given the expected bandwidth $\Delta \omega/\omega \simeq 10^{-6}$, the intrinsic bandwidth of the resonant circuit should be set to $\Delta \omega_\text{LC} = {\rm max}[\Delta \omega, 2\pi/\Delta t]$, where $\Delta t$ is the interrogation time at this frequency.  While $Q_0 \simeq 10^6$ is optimal for sufficiently large $\omega$, smaller 
$Q$ values are needed at smaller $\omega$ to make sure the bandwidth matches the interrogation time.  For example, in the strategy of Ref.~\cite{Chaudhuri:2014dla} where each $e$-fold of frequency is scanned for a time period $t_{\rm e-fold}$, and thus $\Delta t = t_{\rm e-fold}/Q_0$, one must take $Q_0 = \min [10^6, \sqrt{\omega \, t_{\rm e-fold}/2\pi}]$.  Decreasing $Q_0$, however, means adding additional resistance to the circuit and thereby increasing the thermal noise.

Alternatively, we can employ the feedback damping circuit of Refs.~\cite{seton1995use,seton1999gradiometer}, which can widen the intrinsic bandwidth of the resonant circuit without introducing additional noise.  This allows a large $Q$ factor at all frequencies while still capturing all of the signal \cite{myers2007calculated}. The intrinsic $Q_0$ of a niobium superconducting $LC$ circuit is over $10^6$, so we assume $Q_0 = 10^6$ as our benchmark, though larger $Q_0$ may be possible.  The signal flux through the SQUID depends sensitively on the details of the feedback circuit, but our signal-to-noise analysis will not depend on those details, so we treat the feedback circuit as a black box with some inductive coupling $M$ to the SQUID in \Fig{fig:circuits} (right). 

For $Q_0$ up to $\sim$$10^8$, thermal noise in the pickup loop dominates over magnetometer noise (see related studies in Refs.~\cite{Chaudhuri:2014dla,bonaldi1999thermal} and further discussion in the Supplementary Material).  Once we know that thermal noise is dominant, we can calculate the signal-to-noise ratio without regard to the identity of the black box. Following Ref.~\cite{Chaudhuri:2014dla}, the axion sensitivity is set by requiring the signal power dissipated in the resonant circuit to be greater than that of the noise.  The predicted constraints on $g_{a \gamma \gamma}$ depend on how much time is spent on each frequency band.  We imagine a strategy similar to Ref.~\cite{Chaudhuri:2014dla} where each $e$-fold of frequency is scanned for a time period $t_{\rm e-fold}$. To compare with the broadband circuit, we take $t_{\rm e-fold} = 20$ days to cover the frequency range between 1 Hz ($m_a = 4 \times 10^{-15} \ \eV$) and 100 MHz ($m_a = 4 \times 10^{-7} \ \eV$) in the same integration time of 1 year.

At frequency $m_a$, the signal and noise powers are 
\es{powers}{
P_S  = Q_0 \frac{m_a \Phi_{\rm pickup}^2}{2L_T}, \quad P_N = k_B T \sqrt{ \frac{m_a}{2\pi t_{\rm e-fold}}},
}
where $L_T = L_p + L_i$ is the total inductance of the resonant circuit. To compare with the broadband reach we assume $L_T = L_{\rm min}$ as in Eq.~(\ref{eq:Lmin}) and take $h = 3R$.
Requiring a signal-to-noise ratio of unity implies sensitivity to
\es{eq:ResonantReach}{
&\g > 9.0 \times 10^{-17} \ \GeV^{-1} \left(\frac{10^{-12} \ \eV}{m_a} \frac{20 \ {\rm days}}{t_{\rm e-fold}}  \right)^{1/4}   \\
&\times  \frac{5 \ {\rm T}}{B_{\rm max}} \left(\frac{0.85 \ {\rm m}}{R}\right)^{5/2}  \sqrt{\frac{0.3 \ \GeV/{\rm cm}^3}{\rhoDM}\frac{10^6}{Q_0 } \frac{T}{0.1 \ {\rm K}}},
}
where we have assumed a feedback damping circuit that allows us to keep $Q_0$ fixed at low masses.  At high masses, the feedback damping circuit is not necessary unless $Q_0 > 10^6$ is achievable.

As illustrated in \Fig{fig:reach}, the sensitivity increases at \emph{larger} $m_a$ since the signal power density grows as $m_a$. On the other hand, at small masses the broadband approach has a superior projected reach for the same experimental parameters.  Thus, the resonant and broadband approaches are complementary.

We introduced a new experimental design that is sensitive to ultralight DM with axion-like couplings to electromagnetism in the mass range $m_a \in [10^{-14}, 10^{-6}] \ \eV$.  Most existing axion detection proposals use some kind of resonant enhancement, but we have shown that broadband circuits can have superior sensitivity for lighter axion masses.  This conclusion agrees with previous literature establishing that untuned SQUID magnetometers outperform tuned magnetometers at low frequencies \cite{Clarke:1979aa,myers2007calculated}; this fact has been exploited in, e.g.,\ Refs.~\cite{matlachov2004squid,mcdermott2004microtesla} to detect fT magnetic fields from MRI experiments with biological tissue samples. A concrete experiment would likely proceed in two stages: a broadband search over a large frequency range, followed by a resonant scan at high frequencies and in specific frequency bands if a signal is seen. We expect that a broadband magnetometer could also be relevant for detecting dark photon DM \cite{Chaudhuri:2014dla}, and we look forward to further applications of broadband techniques to light DM detection.

\acknowledgments{
{\it Acknowledgments}---We thank Saptarshi Chaudhuri, Kent Irwin, Jeremy Mardon, Lyman Page, Mike Romalis, and Chris Tully for detailed discussions of experimental considerations. In particular, we thank Chris Tully and Mike Romalis for pointing out that we may reduce the geometric inductance of the pickup loop by using smaller loops in parallel, and we thank Kent Irwin for additionally pointing out that there is a minimal pickup-loop inductance allowed by energy conservation.  We thank Asimina Arvanitaki, Dmitry Budker, Simon Coop, Marat Freytsis, Joe Formaggio, Peter Graham, Chris Hill, David E. Kaplan, Rafael Lang, Mariangela Lisanti, David Pinner, and Surjeet Rajendran for helpful conversations. YK thanks Adam Anderson and Bill Jones for enlightening discussions regarding SQUIDs.  BRS is supported by a Pappalardo Fellowship in Physics at MIT.  The work of JT is supported by the U.S. Department of Energy (DOE) under cooperative research agreement DE-SC-00012567, by the DOE Early Career research program DE-SC-0006389, and by a Sloan Research Fellowship from the Alfred P. Sloan Foundation.}

\appendix

\section{Potential noise sources and experimental details}
\label{app:AdditionalNoise}

In our analysis, we estimated the magnetometer noise as (\ref{eq:ZeroTFluxNoise}) and claimed that it dominated in the broadband circuit. This noise level is only a factor of 2 or so above the theoretically predicted temperature-independent floor from current shot noise \cite{Clarke:1979aa}. The spectral density of shot noise is approximately
\es{ShotCurrent}{
S_{J, 0} = \frac{11}{2} e I_0, \qquad I_0 = \frac{\Phi_0}{2L},
}
where $I_0$ is the critical current per Josephson junction in an ideal SQUID. This translates into a flux noise of
\es{FluxShot}{
S_\Phi^{1/2} = L\,  S_{J, 0}^{1/2} = \sqrt{\frac{11}{8}hL}/\sqrt{{\rm Hz}},
}
where $h$ is Planck's constant. Since the signal and shot noise both scale as $\sqrt{L}$ (see (\ref{eq:SQUIDinputflux})), the signal-to-noise ratio is largely independent of the SQUID parameters. 

In a real implementation of our experimental design, magnetic shielding of the entire apparatus will be necessary to reduce environmental noise.  The thermal motion of electrons in the shielding material will itself cause thermal noise, however, with an amplitude proportional to $1/d$, where $d$ is the distance from the shield \cite{lee2008calculation}.  With a superconducting shield, this effect is expected to be small because the only source of thermal noise comes from the thin layer of normal material at the outside of the shield.  Moreover, a superconducting shield would significantly reduce static fluxes compared to a normal conductor such as copper. With a sufficiently large shield cooled to sufficiently low temperatures, we expect that shielding noise will be subdominant at frequencies above 1 kHz \cite{RomalisPrivate}. As long as the shield dimensions are on the order of the toroid size, the signal flux lines will not be significantly distorted at the center of the toroid, and the signal should be relatively unaffected.

An additional source of noise may arise from the static current creating the toroidal $B$-field.  In the ideal scenario, this current does not source any magnetic flux through the center of the toroid, which is a benefit compared to the geometry studied in Ref.~\cite{Sikivie:2013laa}.  One reason this is beneficial is that large fields may make it difficult to maintain the pickup loop in a superconducting phase.  However, a non-uniform geometry, combined with thermal noise in the toroid, may induce static and time-varying flux through the pickup loop due to a small component of the current which circulates azimuthally.  We expect this source of noise to be subdominant in the kHz-GHz range compared to magnetometer noise (in the broadband circuit) or thermal pickup noise (in the resonant circuit).  One possibility for addressing the fringe fields is to circulate a biasing current in the toroid to cancel any static flux though the pickup loop, but this may itself introduce additional thermal noise. While we neglect these noise sources in our analysis, it is important to carefully consider them in a real implementation of the detector. 

To mitigate the effects of $1/f$ noise below 50 Hz, one could attempt to modulate the flux signal either by modulating the toroidal $B$-field or mechanically modulating the pickup loop. Such modulation would likely lead to additional sources of noise, which must be considered in a practical design. We do not attempt to evaluate the contribution of $1/f$ noise in the resonant circuit, which depends on the details of the SQUID coupling, but note that such a contribution will increase at low temperatures below 50 kHz, potentially negating the advantages of operating at lower temperatures.

\section{Dominance of thermal noise for resonant circuits}
\label{app:noise}

When treating the resonant strategy in the body of this letter, we argued that thermal noise in the pickup loop dominates over magnetometer noise.  Here, we illustrate this observation using the feedback damping circuit of Refs.~\cite{seton1995use,seton1999gradiometer}, which is one example of the black box in \Fig{fig:circuits} (right) that couples the $LC$ resonant circuit to the SQUID magnetometer. The effect of the feedback circuit is to increase the bandwidth---in our case, to $\Delta \omega/\omega = \max[ 10^{-6}, 2 \pi / (\Delta t \, \omega) ]$, where $\Delta t$ is the interrogation time at frequency $\omega$---without increasing the noise. We note that this same conclusion, regarding the dominance of thermal noise, was reached for the case of an inductive shunt circuit with a DC SQUID in \cite{Chaudhuri:2014dla}.  Similarly, Ref.~\cite{Chaudhuri:2014dla} considered an AC SQUID readout above 10 MHz, where the SQUID was biased by a microwave-frequency source in order to maintain a sufficiently large $Q$, and thermal noise was dominant in that case as well.   Experimentally, the dominance of thermal noise has been demonstrated for $Q$ up to $10^6$ and $T$ down to 1.2 K for a mechanical-electrical resonator designed to detect gravitational waves \cite{bonaldi1999thermal}.

For the feedback damping circuit, it is useful to generalize \Fig{fig:circuits} to allow the input pickup loop, with inductance $L_p$, to be inductively coupled to an $LC$ circuit, whose inductor has inductance $N_s^2 L_p$.  The total inductance of the $LC$ circuit, including the coupling to the SQUID, is approximately $L_T \approx N_s^2 L_p + L_i$. The separation of the pickup loop from the resonant circuit is useful because, as we will see below and in particular at low frequencies, the optimal $N_s$ may be quite large, in order to minimize thermal noise, while the optimal choice of $L_p$ is always the smallest possible as allowed by energy conservation. Separating the $LC$ circuit from the pickup loop might help mitigate parasitic capacitances. See \cite{Chaudhuri:2014dla} for more details.

In this circuit, the power spectral density of flux noise through the SQUID, $S_\Phi(f)$, at frequency $f$ contains three contributions \cite{myers2007calculated}, namely thermal noise,
\be
\label{eq:ResThermalNoise}
S_\Phi^{T}(f) = \frac{4 k_B T L_T}{N_s^2 \omega Q_0},
\ee
SQUID voltage noise,
\be
\label{eq:ResVoltageNoise}
S_\Phi^V(f) \approx \frac{L_T^2 (\Delta \omega)^2}{N_s^2 \omega^2 M^2 V_\Phi^2}S_V(f),
\ee
and SQUID current noise,
\be
\label{eq:ResCurrentNoise}
S_\Phi^J(f) \approx \frac{M^2}{N_s^2}S_J(f).
\ee
Here, $M = \alpha^2 N_i L$ is the coupling of the input inductor $L_i$ to the SQUID, $\omega$ is the resonant frequency, $\Delta \omega$ is the bandwidth of the resonant circuit including the contribution of feedback damping, and $Q_0 = (\omega C R)^{-1}$ is the intrinsic quality factor of the capacitor.  $V_\Phi$ characterizes the voltage response of the SQUID to a change in flux, and it is roughly expected to be $R/L\sim 10^9 -10^{10} \ {\rm s}^{-1}$.  For a SQUID of junction resistance $R$, $S_V(f) \approx 16 k_B T R$ and \mbox{$S_J(f) \approx 11 k_B T/R + S_{J, 0}$} are the spectral densities of voltage and current noise  \cite{Clarke:1979aa}.  We have explicitly added the irreducible shot noise contribution $S_{J, 0}$ to the SQUID current noise (see (\ref{ShotCurrent})); this term is negligible at high temperatures but begins to dominate below $\sim$60 mK.

The optimal number of turns $N_s$ is determined by minimizing the magnetic flux noise through the SQUID with respect to $N_s$~\cite{myers2007calculated}, yielding
\es{eq:OptNp}{
N_s^2 L_p = &L_i  \left[1 + \frac{\alpha^2 \omega V_\Phi}{4Q_0 (\Delta \omega)^2} \right. \\
 &\left.+ \frac{\alpha^4 R \omega^2}{16 k_B T (\Delta \omega)^2} 
\left(\frac{11 k_B T}{R} + S_{J, 0} \right) \right]^{1/2} \,.
}
For $\Delta \omega/\omega = 10^{-6}$, $\alpha^2 = 0.5$, frequencies $\omega < 10 \ {\rm MHz}$, and $Q_0 > 10^6$, the last term in \eqref{eq:OptNp} dominates, giving $N_s^2 L_p \approx 10^6 L_i$ and thus $L_T \approx 10^6 L_i$. Depending on the maximum attainable capacitance, the optimal $N_s$ may be quite large at low frequencies. As an example, Ref.~\cite{myers2007calculated} estimates a maximum low-loss capacitance of 0.1 $\mu{\rm F}$, such that $N_s \simeq 10^5$ for $\omega = 10 \ {\rm Hz}$, given $L_p$ as calculated in \eqref{eq:PickupLoopL}. 

Substituting \eqref{eq:OptNp} into \eqref{eq:ResThermalNoise}--\eqref{eq:ResCurrentNoise} gives
\begin{align}
\label{eq:OptResNoise}
S_\Phi(f) \approx \frac{4 k_B T L_p}{ \omega Q_0} & \left [ 1  + \frac{4 \times 10^{-6}}{\alpha^2} \frac{Q_0 \Delta \omega}{\omega} \frac{\Delta\omega}{V_\Phi} \right. \\ \nonumber
& \left. ~+ 10^{-6} Q_0 \alpha^2 \frac{\omega}{V_\Phi} \left (\frac{11}{4} + \frac{S_{J, 0}}{4 k_B T} \right) \right ],
\end{align}
where the three terms correspond to thermal noise, SQUID voltage noise, and SQUID current noise. For the parameters of interest, the second term is always subdominant to the third term. Since $\omega/V_\Phi \lesssim 10^{-2}$, the third term is suppressed compared to the first for $Q_0 \lesssim 10^8$. As discussed in Ref.~\cite{Chaudhuri:2014dla}, $Q_0$ for a niobium superconducting $LC$ circuit is at least $10^6$, but achieving $Q_0$ of $10^8$ is difficult. Thus, thermal noise in the $LC$ resonant circuit dominates the flux noise in the tuned magnetometer below 100 MHz, as anticipated.

It is useful to make contact with the untuned magnetometer in this framework.  Ignoring for the moment the finite bandwidth of the signal, as we imagine taking \mbox{$Q_0 \to \infty$} at fixed $L$ and $C$, the resistance in the resonant circuit disappears and magnetometer noise should dominate. Indeed, in that limit the first term in \eqref{eq:OptResNoise} is suppressed, leaving dominantly the current noise, as we found for the broadband circuit:
\be
\left. S_\Phi(f) \right|_{Q_0\to \infty}  = (11 k_B T + S_{J, 0})M^2/ N_s^2 \,.
\ee
Note that this equation refers to flux noise through the SQUID, and \eqref{eq:SQUIDinputflux} can be used to determine the input flux noise. Also note that $M^2 \propto N_i^2 \propto N_s^2$, so that $S_\Phi$ is independent of $N_s$ in the broadband case. 

\bibliography{Axion_BroadbandBib}

\begin{thebibliography}{54}%
\makeatletter
\providecommand \@ifxundefined [1]{%
 \@ifx{#1\undefined}
}%
\providecommand \@ifnum [1]{%
 \ifnum #1\expandafter \@firstoftwo
 \else \expandafter \@secondoftwo
 \fi
}%
\providecommand \@ifx [1]{%
 \ifx #1\expandafter \@firstoftwo
 \else \expandafter \@secondoftwo
 \fi
}%
\providecommand \natexlab [1]{#1}%
\providecommand \enquote  [1]{``#1''}%
\providecommand \bibnamefont  [1]{#1}%
\providecommand \bibfnamefont [1]{#1}%
\providecommand \citenamefont [1]{#1}%
\providecommand \href@noop [0]{\@secondoftwo}%
\providecommand \href [0]{\begingroup \@sanitize@url \@href}%
\providecommand \@href[1]{\@@startlink{#1}\@@href}%
\providecommand \@@href[1]{\endgroup#1\@@endlink}%
\providecommand \@sanitize@url [0]{\catcode `\\12\catcode `\$12\catcode
  `\&12\catcode `\#12\catcode `\^12\catcode `\_12\catcode `\%12\relax}%
\providecommand \@@startlink[1]{}%
\providecommand \@@endlink[0]{}%
\providecommand \url  [0]{\begingroup\@sanitize@url \@url }%
\providecommand \@url [1]{\endgroup\@href {#1}{\urlprefix }}%
\providecommand \urlprefix  [0]{URL }%
\providecommand \Eprint [0]{\href }%
\providecommand \doibase [0]{http://dx.doi.org/}%
\providecommand \selectlanguage [0]{\@gobble}%
\providecommand \bibinfo  [0]{\@secondoftwo}%
\providecommand \bibfield  [0]{\@secondoftwo}%
\providecommand \translation [1]{[#1]}%
\providecommand \BibitemOpen [0]{}%
\providecommand \bibitemStop [0]{}%
\providecommand \bibitemNoStop [0]{.\EOS\space}%
\providecommand \EOS [0]{\spacefactor3000\relax}%
\providecommand \BibitemShut  [1]{\csname bibitem#1\endcsname}%
\let\auto@bib@innerbib\@empty
\bibitem [{\citenamefont {Preskill}\ \emph {et~al.}(1983)\citenamefont
  {Preskill}, \citenamefont {Wise},\ and\ \citenamefont
  {Wilczek}}]{Preskill:1982cy}%
  \BibitemOpen
  \bibfield  {author} {\bibinfo {author} {\bibfnamefont {John}\ \bibnamefont
  {Preskill}}, \bibinfo {author} {\bibfnamefont {Mark~B.}\ \bibnamefont
  {Wise}}, \ and\ \bibinfo {author} {\bibfnamefont {Frank}\ \bibnamefont
  {Wilczek}},\ }\bibfield  {title} {\enquote {\bibinfo {title} {{Cosmology of
  the Invisible Axion}},}\ }\href {\doibase 10.1016/0370-2693(83)90637-8}
  {\bibfield  {journal} {\bibinfo  {journal} {Phys. Lett.}\ }\textbf {\bibinfo
  {volume} {B120}},\ \bibinfo {pages} {127--132} (\bibinfo {year}
  {1983})}\BibitemShut {NoStop}%
\bibitem [{\citenamefont {Abbott}\ and\ \citenamefont
  {Sikivie}(1983)}]{Abbott:1982af}%
  \BibitemOpen
  \bibfield  {author} {\bibinfo {author} {\bibfnamefont {L.~F.}\ \bibnamefont
  {Abbott}}\ and\ \bibinfo {author} {\bibfnamefont {P.}~\bibnamefont
  {Sikivie}},\ }\bibfield  {title} {\enquote {\bibinfo {title} {{A Cosmological
  Bound on the Invisible Axion}},}\ }\href {\doibase
  10.1016/0370-2693(83)90638-X} {\bibfield  {journal} {\bibinfo  {journal}
  {Phys. Lett.}\ }\textbf {\bibinfo {volume} {B120}},\ \bibinfo {pages}
  {133--136} (\bibinfo {year} {1983})}\BibitemShut {NoStop}%
\bibitem [{\citenamefont {Dine}\ and\ \citenamefont
  {Fischler}(1983)}]{Dine:1982ah}%
  \BibitemOpen
  \bibfield  {author} {\bibinfo {author} {\bibfnamefont {Michael}\ \bibnamefont
  {Dine}}\ and\ \bibinfo {author} {\bibfnamefont {Willy}\ \bibnamefont
  {Fischler}},\ }\bibfield  {title} {\enquote {\bibinfo {title} {{The Not So
  Harmless Axion}},}\ }\href {\doibase 10.1016/0370-2693(83)90639-1} {\bibfield
   {journal} {\bibinfo  {journal} {Phys. Lett.}\ }\textbf {\bibinfo {volume}
  {B120}},\ \bibinfo {pages} {137--141} (\bibinfo {year} {1983})}\BibitemShut
  {NoStop}%
\bibitem [{\citenamefont {Peccei}\ and\ \citenamefont
  {Quinn}(1977{\natexlab{a}})}]{Peccei:1977hh}%
  \BibitemOpen
  \bibfield  {author} {\bibinfo {author} {\bibfnamefont {R.~D.}\ \bibnamefont
  {Peccei}}\ and\ \bibinfo {author} {\bibfnamefont {Helen~R.}\ \bibnamefont
  {Quinn}},\ }\bibfield  {title} {\enquote {\bibinfo {title} {{CP Conservation
  in the Presence of Instantons}},}\ }\href {\doibase
  10.1103/PhysRevLett.38.1440} {\bibfield  {journal} {\bibinfo  {journal}
  {Phys. Rev. Lett.}\ }\textbf {\bibinfo {volume} {38}},\ \bibinfo {pages}
  {1440--1443} (\bibinfo {year} {1977}{\natexlab{a}})}\BibitemShut {NoStop}%
\bibitem [{\citenamefont {Peccei}\ and\ \citenamefont
  {Quinn}(1977{\natexlab{b}})}]{Peccei:1977ur}%
  \BibitemOpen
  \bibfield  {author} {\bibinfo {author} {\bibfnamefont {R.~D.}\ \bibnamefont
  {Peccei}}\ and\ \bibinfo {author} {\bibfnamefont {Helen~R.}\ \bibnamefont
  {Quinn}},\ }\bibfield  {title} {\enquote {\bibinfo {title} {{Constraints
  Imposed by CP Conservation in the Presence of Instantons}},}\ }\href
  {\doibase 10.1103/PhysRevD.16.1791} {\bibfield  {journal} {\bibinfo
  {journal} {Phys. Rev.}\ }\textbf {\bibinfo {volume} {D16}},\ \bibinfo {pages}
  {1791--1797} (\bibinfo {year} {1977}{\natexlab{b}})}\BibitemShut {NoStop}%
\bibitem [{\citenamefont {Weinberg}(1978)}]{Weinberg:1977ma}%
  \BibitemOpen
  \bibfield  {author} {\bibinfo {author} {\bibfnamefont {Steven}\ \bibnamefont
  {Weinberg}},\ }\bibfield  {title} {\enquote {\bibinfo {title} {{A New Light
  Boson?}}}\ }\href {\doibase 10.1103/PhysRevLett.40.223} {\bibfield  {journal}
  {\bibinfo  {journal} {Phys.Rev.Lett.}\ }\textbf {\bibinfo {volume} {40}},\
  \bibinfo {pages} {223--226} (\bibinfo {year} {1978})}\BibitemShut {NoStop}%
\bibitem [{\citenamefont {Wilczek}(1978)}]{Wilczek:1977pj}%
  \BibitemOpen
  \bibfield  {author} {\bibinfo {author} {\bibfnamefont {Frank}\ \bibnamefont
  {Wilczek}},\ }\bibfield  {title} {\enquote {\bibinfo {title} {{Problem of
  Strong P and T Invariance in the Presence of Instantons}},}\ }\href {\doibase
  10.1103/PhysRevLett.40.279} {\bibfield  {journal} {\bibinfo  {journal}
  {Phys.Rev.Lett.}\ }\textbf {\bibinfo {volume} {40}},\ \bibinfo {pages}
  {279--282} (\bibinfo {year} {1978})}\BibitemShut {NoStop}%
\bibitem [{\citenamefont {Svrcek}\ and\ \citenamefont
  {Witten}(2006)}]{Svrcek:2006yi}%
  \BibitemOpen
  \bibfield  {author} {\bibinfo {author} {\bibfnamefont {Peter}\ \bibnamefont
  {Svrcek}}\ and\ \bibinfo {author} {\bibfnamefont {Edward}\ \bibnamefont
  {Witten}},\ }\bibfield  {title} {\enquote {\bibinfo {title} {{Axions In
  String Theory}},}\ }\href {\doibase 10.1088/1126-6708/2006/06/051} {\bibfield
   {journal} {\bibinfo  {journal} {JHEP}\ }\textbf {\bibinfo {volume} {06}},\
  \bibinfo {pages} {051} (\bibinfo {year} {2006})},\ \Eprint
  {http://arxiv.org/abs/hep-th/0605206} {arXiv:hep-th/0605206 [hep-th]}
  \BibitemShut {NoStop}%
\bibitem [{\citenamefont {Essig}\ \emph {et~al.}(2013)\citenamefont {Essig}
  \emph {et~al.}}]{Essig:2013lka}%
  \BibitemOpen
  \bibfield  {author} {\bibinfo {author} {\bibfnamefont {Rouven}\ \bibnamefont
  {Essig}} \emph {et~al.},\ }\bibfield  {title} {\enquote {\bibinfo {title}
  {{Working Group Report: New Light Weakly Coupled Particles}},}\ }in\ \href
  {http://inspirehep.net/record/1263039/files/arXiv:1311.0029.pdf} {\emph
  {\bibinfo {booktitle} {{Community Summer Study 2013: Snowmass on the
  Mississippi (CSS2013) Minneapolis, MN, USA, July 29-August 6, 2013}}}}\
  (\bibinfo {year} {2013})\ \Eprint {http://arxiv.org/abs/1311.0029}
  {arXiv:1311.0029 [hep-ph]} \BibitemShut {NoStop}%
\bibitem [{\citenamefont {Marsh}(2015)}]{Marsh:2015xka}%
  \BibitemOpen
  \bibfield  {author} {\bibinfo {author} {\bibfnamefont {David J.~E.}\
  \bibnamefont {Marsh}},\ }\bibfield  {title} {\enquote {\bibinfo {title}
  {{Axion Cosmology}},}\ }\href@noop {} {\  (\bibinfo {year} {2015})},\ \Eprint
  {http://arxiv.org/abs/1510.07633} {arXiv:1510.07633 [astro-ph.CO]}
  \BibitemShut {NoStop}%
\bibitem [{\citenamefont {Graham}\ \emph {et~al.}(2015)\citenamefont {Graham},
  \citenamefont {Irastorza}, \citenamefont {Lamoreaux}, \citenamefont
  {Lindner},\ and\ \citenamefont {van Bibber}}]{Graham:2015ouw}%
  \BibitemOpen
  \bibfield  {author} {\bibinfo {author} {\bibfnamefont {Peter~W.}\
  \bibnamefont {Graham}}, \bibinfo {author} {\bibfnamefont {Igor~G.}\
  \bibnamefont {Irastorza}}, \bibinfo {author} {\bibfnamefont {Steven~K.}\
  \bibnamefont {Lamoreaux}}, \bibinfo {author} {\bibfnamefont {Axel}\
  \bibnamefont {Lindner}}, \ and\ \bibinfo {author} {\bibfnamefont {Karl~A.}\
  \bibnamefont {van Bibber}},\ }\bibfield  {title} {\enquote {\bibinfo {title}
  {{Experimental Searches for the Axion and Axion-Like Particles}},}\ }\href
  {\doibase 10.1146/annurev-nucl-102014-022120} {\bibfield  {journal} {\bibinfo
   {journal} {Ann. Rev. Nucl. Part. Sci.}\ }\textbf {\bibinfo {volume} {65}},\
  \bibinfo {pages} {485--514} (\bibinfo {year} {2015})},\ \Eprint
  {http://arxiv.org/abs/1602.00039} {arXiv:1602.00039 [hep-ex]} \BibitemShut
  {NoStop}%
\bibitem [{\citenamefont {Asztalos}\ \emph {et~al.}(2001)\citenamefont
  {Asztalos} \emph {et~al.}}]{Asztalos:2001tf}%
  \BibitemOpen
  \bibfield  {author} {\bibinfo {author} {\bibfnamefont {Stephen~J.}\
  \bibnamefont {Asztalos}} \emph {et~al.} (\bibinfo {collaboration} {ADMX}),\
  }\bibfield  {title} {\enquote {\bibinfo {title} {{Large scale microwave
  cavity search for dark matter axions}},}\ }\href {\doibase
  10.1103/PhysRevD.64.092003} {\bibfield  {journal} {\bibinfo  {journal} {Phys.
  Rev.}\ }\textbf {\bibinfo {volume} {D64}},\ \bibinfo {pages} {092003}
  (\bibinfo {year} {2001})}\BibitemShut {NoStop}%
\bibitem [{\citenamefont {Asztalos}\ \emph {et~al.}(2010)\citenamefont
  {Asztalos}, \citenamefont {Carosi}, \citenamefont {Hagmann}, \citenamefont
  {Kinion}, \citenamefont {van Bibber}, \citenamefont {Hotz}, \citenamefont
  {Rosenberg}, \citenamefont {Rybka}, \citenamefont {Hoskins}, \citenamefont
  {Hwang}, \citenamefont {Sikivie}, \citenamefont {Tanner}, \citenamefont
  {Bradley},\ and\ \citenamefont {Clarke}}]{PhysRevLett.104.041301}%
  \BibitemOpen
  \bibfield  {author} {\bibinfo {author} {\bibfnamefont {S.~J.}\ \bibnamefont
  {Asztalos}}, \bibinfo {author} {\bibfnamefont {G.}~\bibnamefont {Carosi}},
  \bibinfo {author} {\bibfnamefont {C.}~\bibnamefont {Hagmann}}, \bibinfo
  {author} {\bibfnamefont {D.}~\bibnamefont {Kinion}}, \bibinfo {author}
  {\bibfnamefont {K.}~\bibnamefont {van Bibber}}, \bibinfo {author}
  {\bibfnamefont {M.}~\bibnamefont {Hotz}}, \bibinfo {author} {\bibfnamefont
  {L.~J}\ \bibnamefont {Rosenberg}}, \bibinfo {author} {\bibfnamefont
  {G.}~\bibnamefont {Rybka}}, \bibinfo {author} {\bibfnamefont
  {J.}~\bibnamefont {Hoskins}}, \bibinfo {author} {\bibfnamefont
  {J.}~\bibnamefont {Hwang}}, \bibinfo {author} {\bibfnamefont
  {P.}~\bibnamefont {Sikivie}}, \bibinfo {author} {\bibfnamefont {D.~B.}\
  \bibnamefont {Tanner}}, \bibinfo {author} {\bibfnamefont {R.}~\bibnamefont
  {Bradley}}, \ and\ \bibinfo {author} {\bibfnamefont {J.}~\bibnamefont
  {Clarke}},\ }\bibfield  {title} {\enquote {\bibinfo {title} {{SQUID-Based
  Microwave Cavity Search for Dark-Matter Axions}},}\ }\href {\doibase
  10.1103/PhysRevLett.104.041301} {\bibfield  {journal} {\bibinfo  {journal}
  {Phys. Rev. Lett.}\ }\textbf {\bibinfo {volume} {104}},\ \bibinfo {pages}
  {041301} (\bibinfo {year} {2010})}\BibitemShut {NoStop}%
\bibitem [{\citenamefont {Shokair}\ \emph {et~al.}(2014)\citenamefont {Shokair}
  \emph {et~al.}}]{Shokair:2014rna}%
  \BibitemOpen
  \bibfield  {author} {\bibinfo {author} {\bibfnamefont {T.~M.}\ \bibnamefont
  {Shokair}} \emph {et~al.},\ }\bibfield  {title} {\enquote {\bibinfo {title}
  {{Future Directions in the Microwave Cavity Search for Dark Matter
  Axions}},}\ }\href {\doibase 10.1142/S0217751X14430040} {\bibfield  {journal}
  {\bibinfo  {journal} {Int. J. Mod. Phys.}\ }\textbf {\bibinfo {volume}
  {A29}},\ \bibinfo {pages} {1443004} (\bibinfo {year} {2014})},\ \Eprint
  {http://arxiv.org/abs/1405.3685} {arXiv:1405.3685 [physics.ins-det]}
  \BibitemShut {NoStop}%
\bibitem [{\citenamefont {Read}(2014)}]{Read:2014qva}%
  \BibitemOpen
  \bibfield  {author} {\bibinfo {author} {\bibfnamefont {J.~I.}\ \bibnamefont
  {Read}},\ }\bibfield  {title} {\enquote {\bibinfo {title} {{The Local Dark
  Matter Density}},}\ }\href {\doibase 10.1088/0954-3899/41/6/063101}
  {\bibfield  {journal} {\bibinfo  {journal} {J. Phys.}\ }\textbf {\bibinfo
  {volume} {G41}},\ \bibinfo {pages} {063101} (\bibinfo {year} {2014})},\
  \Eprint {http://arxiv.org/abs/1404.1938} {arXiv:1404.1938 [astro-ph.GA]}
  \BibitemShut {NoStop}%
\bibitem [{\citenamefont {Sikivie}(1983)}]{Sikivie:1983ip}%
  \BibitemOpen
  \bibfield  {author} {\bibinfo {author} {\bibfnamefont {P.}~\bibnamefont
  {Sikivie}},\ }\bibfield  {title} {\enquote {\bibinfo {title} {{Experimental
  Tests of the Invisible Axion}},}\ }\bibfield  {booktitle} {\emph {\bibinfo
  {booktitle} {{11th International Symposium on Lepton and Photon Interactions
  at High Energies Ithaca, New York, August 4-9, 1983}}},\ }\href {\doibase
  10.1103/PhysRevLett.51.1415} {\bibfield  {journal} {\bibinfo  {journal}
  {Phys. Rev. Lett.}\ }\textbf {\bibinfo {volume} {51}},\ \bibinfo {pages}
  {1415--1417} (\bibinfo {year} {1983})},\ \bibinfo {note} {[Erratum: Phys.
  Rev. Lett.52,695(1984)]}\BibitemShut {NoStop}%
\bibitem [{\citenamefont {Dine}\ \emph {et~al.}(1981)\citenamefont {Dine},
  \citenamefont {Fischler},\ and\ \citenamefont {Srednicki}}]{Dine:1981rt}%
  \BibitemOpen
  \bibfield  {author} {\bibinfo {author} {\bibfnamefont {Michael}\ \bibnamefont
  {Dine}}, \bibinfo {author} {\bibfnamefont {Willy}\ \bibnamefont {Fischler}},
  \ and\ \bibinfo {author} {\bibfnamefont {Mark}\ \bibnamefont {Srednicki}},\
  }\bibfield  {title} {\enquote {\bibinfo {title} {{A Simple Solution to the
  Strong CP Problem with a Harmless Axion}},}\ }\href {\doibase
  10.1016/0370-2693(81)90590-6} {\bibfield  {journal} {\bibinfo  {journal}
  {Phys. Lett.}\ }\textbf {\bibinfo {volume} {B104}},\ \bibinfo {pages} {199}
  (\bibinfo {year} {1981})}\BibitemShut {NoStop}%
\bibitem [{\citenamefont {Zhitnitsky}(1980)}]{Zhitnitsky:1980tq}%
  \BibitemOpen
  \bibfield  {author} {\bibinfo {author} {\bibfnamefont {A.~R.}\ \bibnamefont
  {Zhitnitsky}},\ }\bibfield  {title} {\enquote {\bibinfo {title} {{On Possible
  Suppression of the Axion Hadron Interactions. (In Russian)}},}\ }\href@noop
  {} {\bibfield  {journal} {\bibinfo  {journal} {Sov. J. Nucl. Phys.}\ }\textbf
  {\bibinfo {volume} {31}},\ \bibinfo {pages} {260} (\bibinfo {year} {1980})},\
  \bibinfo {note} {[Yad. Fiz.31,497(1980)]}\BibitemShut {NoStop}%
\bibitem [{\citenamefont {Kim}(1979)}]{Kim:1979if}%
  \BibitemOpen
  \bibfield  {author} {\bibinfo {author} {\bibfnamefont {Jihn~E.}\ \bibnamefont
  {Kim}},\ }\bibfield  {title} {\enquote {\bibinfo {title} {{Weak Interaction
  Singlet and Strong CP Invariance}},}\ }\href {\doibase
  10.1103/PhysRevLett.43.103} {\bibfield  {journal} {\bibinfo  {journal} {Phys.
  Rev. Lett.}\ }\textbf {\bibinfo {volume} {43}},\ \bibinfo {pages} {103}
  (\bibinfo {year} {1979})}\BibitemShut {NoStop}%
\bibitem [{\citenamefont {Shifman}\ \emph {et~al.}(1980)\citenamefont
  {Shifman}, \citenamefont {Vainshtein},\ and\ \citenamefont
  {Zakharov}}]{Shifman:1979if}%
  \BibitemOpen
  \bibfield  {author} {\bibinfo {author} {\bibfnamefont {Mikhail~A.}\
  \bibnamefont {Shifman}}, \bibinfo {author} {\bibfnamefont {A.~I.}\
  \bibnamefont {Vainshtein}}, \ and\ \bibinfo {author} {\bibfnamefont
  {Valentin~I.}\ \bibnamefont {Zakharov}},\ }\bibfield  {title} {\enquote
  {\bibinfo {title} {{Can Confinement Ensure Natural CP Invariance of Strong
  Interactions?}}}\ }\href {\doibase 10.1016/0550-3213(80)90209-6} {\bibfield
  {journal} {\bibinfo  {journal} {Nucl. Phys.}\ }\textbf {\bibinfo {volume}
  {B166}},\ \bibinfo {pages} {493} (\bibinfo {year} {1980})}\BibitemShut
  {NoStop}%
\bibitem [{\citenamefont {Grohmann}\ \emph {et~al.}(1974)\citenamefont
  {Grohmann}, \citenamefont {Hahlbohm}, \citenamefont {L{\"u}bbig},\ and\
  \citenamefont {Ramin}}]{grohmann1974current}%
  \BibitemOpen
  \bibfield  {author} {\bibinfo {author} {\bibfnamefont {K}~\bibnamefont
  {Grohmann}}, \bibinfo {author} {\bibfnamefont {HD}~\bibnamefont {Hahlbohm}},
  \bibinfo {author} {\bibfnamefont {H}~\bibnamefont {L{\"u}bbig}}, \ and\
  \bibinfo {author} {\bibfnamefont {H}~\bibnamefont {Ramin}},\ }\bibfield
  {title} {\enquote {\bibinfo {title} {Current comparators with superconducting
  shields},}\ }\href@noop {} {\bibfield  {journal} {\bibinfo  {journal}
  {Cryogenics}\ }\textbf {\bibinfo {volume} {14}},\ \bibinfo {pages} {499--502}
  (\bibinfo {year} {1974})}\BibitemShut {NoStop}%
\bibitem [{\citenamefont {Sikivie}\ \emph {et~al.}(2014)\citenamefont
  {Sikivie}, \citenamefont {Sullivan},\ and\ \citenamefont
  {Tanner}}]{Sikivie:2013laa}%
  \BibitemOpen
  \bibfield  {author} {\bibinfo {author} {\bibfnamefont {P.}~\bibnamefont
  {Sikivie}}, \bibinfo {author} {\bibfnamefont {N.}~\bibnamefont {Sullivan}}, \
  and\ \bibinfo {author} {\bibfnamefont {D.~B.}\ \bibnamefont {Tanner}},\
  }\bibfield  {title} {\enquote {\bibinfo {title} {{Proposal for Axion Dark
  Matter Detection Using an LC Circuit}},}\ }\href {\doibase
  10.1103/PhysRevLett.112.131301} {\bibfield  {journal} {\bibinfo  {journal}
  {Phys. Rev. Lett.}\ }\textbf {\bibinfo {volume} {112}},\ \bibinfo {pages}
  {131301} (\bibinfo {year} {2014})},\ \Eprint {http://arxiv.org/abs/1310.8545}
  {arXiv:1310.8545 [hep-ph]} \BibitemShut {NoStop}%
\bibitem [{\citenamefont {Thomas}\ and\ \citenamefont
  {Cabrera}()}]{ThomasTalk}%
  \BibitemOpen
  \bibfield  {author} {\bibinfo {author} {\bibfnamefont {Scott}\ \bibnamefont
  {Thomas}}\ and\ \bibinfo {author} {\bibfnamefont {Blas}\ \bibnamefont
  {Cabrera}},\ }\href@noop {} {\enquote {\bibinfo {title} {Detecting
  string-scale {QCD} axion dark matter},}\ }\bibinfo {howpublished} {Conference
  talk at Axions 2010.}\BibitemShut {Stop}%
\bibitem [{\citenamefont {Chaudhuri}\ \emph {et~al.}(2015)\citenamefont
  {Chaudhuri}, \citenamefont {Graham}, \citenamefont {Irwin}, \citenamefont
  {Mardon}, \citenamefont {Rajendran},\ and\ \citenamefont
  {Zhao}}]{Chaudhuri:2014dla}%
  \BibitemOpen
  \bibfield  {author} {\bibinfo {author} {\bibfnamefont {Saptarshi}\
  \bibnamefont {Chaudhuri}}, \bibinfo {author} {\bibfnamefont {Peter~W.}\
  \bibnamefont {Graham}}, \bibinfo {author} {\bibfnamefont {Kent}\ \bibnamefont
  {Irwin}}, \bibinfo {author} {\bibfnamefont {Jeremy}\ \bibnamefont {Mardon}},
  \bibinfo {author} {\bibfnamefont {Surjeet}\ \bibnamefont {Rajendran}}, \ and\
  \bibinfo {author} {\bibfnamefont {Yue}\ \bibnamefont {Zhao}},\ }\bibfield
  {title} {\enquote {\bibinfo {title} {{Radio for hidden-photon dark matter
  detection}},}\ }\href {\doibase 10.1103/PhysRevD.92.075012} {\bibfield
  {journal} {\bibinfo  {journal} {Phys. Rev.}\ }\textbf {\bibinfo {volume}
  {D92}},\ \bibinfo {pages} {075012} (\bibinfo {year} {2015})},\ \Eprint
  {http://arxiv.org/abs/1411.7382} {arXiv:1411.7382 [hep-ph]} \BibitemShut
  {NoStop}%
\bibitem [{\citenamefont {Budker}\ \emph {et~al.}(2014)\citenamefont {Budker},
  \citenamefont {Graham}, \citenamefont {Ledbetter}, \citenamefont
  {Rajendran},\ and\ \citenamefont {Sushkov}}]{Budker:2013hfa}%
  \BibitemOpen
  \bibfield  {author} {\bibinfo {author} {\bibfnamefont {Dmitry}\ \bibnamefont
  {Budker}}, \bibinfo {author} {\bibfnamefont {Peter~W.}\ \bibnamefont
  {Graham}}, \bibinfo {author} {\bibfnamefont {Micah}\ \bibnamefont
  {Ledbetter}}, \bibinfo {author} {\bibfnamefont {Surjeet}\ \bibnamefont
  {Rajendran}}, \ and\ \bibinfo {author} {\bibfnamefont {Alex}\ \bibnamefont
  {Sushkov}},\ }\bibfield  {title} {\enquote {\bibinfo {title} {{Proposal for a
  Cosmic Axion Spin Precession Experiment (CASPEr)}},}\ }\href {\doibase
  10.1103/PhysRevX.4.021030} {\bibfield  {journal} {\bibinfo  {journal} {Phys.
  Rev.}\ }\textbf {\bibinfo {volume} {X4}},\ \bibinfo {pages} {021030}
  (\bibinfo {year} {2014})},\ \Eprint {http://arxiv.org/abs/1306.6089}
  {arXiv:1306.6089 [hep-ph]} \BibitemShut {NoStop}%
\bibitem [{\citenamefont {Baker}\ \emph {et~al.}(2012)\citenamefont {Baker},
  \citenamefont {Betz}, \citenamefont {Caspers}, \citenamefont {Jaeckel},
  \citenamefont {Lindner}, \citenamefont {Ringwald}, \citenamefont
  {Semertzidis}, \citenamefont {Sikivie},\ and\ \citenamefont
  {Zioutas}}]{Baker:2011na}%
  \BibitemOpen
  \bibfield  {author} {\bibinfo {author} {\bibfnamefont {Oliver~K.}\
  \bibnamefont {Baker}}, \bibinfo {author} {\bibfnamefont {Michael}\
  \bibnamefont {Betz}}, \bibinfo {author} {\bibfnamefont {Fritz}\ \bibnamefont
  {Caspers}}, \bibinfo {author} {\bibfnamefont {Joerg}\ \bibnamefont
  {Jaeckel}}, \bibinfo {author} {\bibfnamefont {Axel}\ \bibnamefont {Lindner}},
  \bibinfo {author} {\bibfnamefont {Andreas}\ \bibnamefont {Ringwald}},
  \bibinfo {author} {\bibfnamefont {Yannis}\ \bibnamefont {Semertzidis}},
  \bibinfo {author} {\bibfnamefont {Pierre}\ \bibnamefont {Sikivie}}, \ and\
  \bibinfo {author} {\bibfnamefont {Konstantin}\ \bibnamefont {Zioutas}},\
  }\bibfield  {title} {\enquote {\bibinfo {title} {{Prospects for Searching
  Axion-like Particle Dark Matter with Dipole, Toroidal and Wiggler
  Magnets}},}\ }\href {\doibase 10.1103/PhysRevD.85.035018} {\bibfield
  {journal} {\bibinfo  {journal} {Phys. Rev.}\ }\textbf {\bibinfo {volume}
  {D85}},\ \bibinfo {pages} {035018} (\bibinfo {year} {2012})},\ \Eprint
  {http://arxiv.org/abs/1110.2180} {arXiv:1110.2180 [physics.ins-det]}
  \BibitemShut {NoStop}%
\bibitem [{\citenamefont {Graham}\ and\ \citenamefont
  {Rajendran}(2011)}]{Graham:2011qk}%
  \BibitemOpen
  \bibfield  {author} {\bibinfo {author} {\bibfnamefont {Peter~W.}\
  \bibnamefont {Graham}}\ and\ \bibinfo {author} {\bibfnamefont {Surjeet}\
  \bibnamefont {Rajendran}},\ }\bibfield  {title} {\enquote {\bibinfo {title}
  {{Axion Dark Matter Detection with Cold Molecules}},}\ }\href {\doibase
  10.1103/PhysRevD.84.055013} {\bibfield  {journal} {\bibinfo  {journal} {Phys.
  Rev.}\ }\textbf {\bibinfo {volume} {D84}},\ \bibinfo {pages} {055013}
  (\bibinfo {year} {2011})},\ \Eprint {http://arxiv.org/abs/1101.2691}
  {arXiv:1101.2691 [hep-ph]} \BibitemShut {NoStop}%
\bibitem [{\citenamefont {Graham}\ and\ \citenamefont
  {Rajendran}(2013)}]{Graham:2013gfa}%
  \BibitemOpen
  \bibfield  {author} {\bibinfo {author} {\bibfnamefont {Peter~W.}\
  \bibnamefont {Graham}}\ and\ \bibinfo {author} {\bibfnamefont {Surjeet}\
  \bibnamefont {Rajendran}},\ }\bibfield  {title} {\enquote {\bibinfo {title}
  {{New Observables for Direct Detection of Axion Dark Matter}},}\ }\href
  {\doibase 10.1103/PhysRevD.88.035023} {\bibfield  {journal} {\bibinfo
  {journal} {Phys. Rev.}\ }\textbf {\bibinfo {volume} {D88}},\ \bibinfo {pages}
  {035023} (\bibinfo {year} {2013})},\ \Eprint {http://arxiv.org/abs/1306.6088}
  {arXiv:1306.6088 [hep-ph]} \BibitemShut {NoStop}%
\bibitem [{\citenamefont {Horns}\ \emph
  {et~al.}(2013{\natexlab{a}})\citenamefont {Horns}, \citenamefont {Jaeckel},
  \citenamefont {Lindner}, \citenamefont {Lobanov}, \citenamefont {Redondo},\
  and\ \citenamefont {Ringwald}}]{Horns:2012jf}%
  \BibitemOpen
  \bibfield  {author} {\bibinfo {author} {\bibfnamefont {Dieter}\ \bibnamefont
  {Horns}}, \bibinfo {author} {\bibfnamefont {Joerg}\ \bibnamefont {Jaeckel}},
  \bibinfo {author} {\bibfnamefont {Axel}\ \bibnamefont {Lindner}}, \bibinfo
  {author} {\bibfnamefont {Andrei}\ \bibnamefont {Lobanov}}, \bibinfo {author}
  {\bibfnamefont {Javier}\ \bibnamefont {Redondo}}, \ and\ \bibinfo {author}
  {\bibfnamefont {Andreas}\ \bibnamefont {Ringwald}},\ }\bibfield  {title}
  {\enquote {\bibinfo {title} {{Searching for WISPy Cold Dark Matter with a
  Dish Antenna}},}\ }\href {\doibase 10.1088/1475-7516/2013/04/016} {\bibfield
  {journal} {\bibinfo  {journal} {JCAP}\ }\textbf {\bibinfo {volume} {1304}},\
  \bibinfo {pages} {016} (\bibinfo {year} {2013}{\natexlab{a}})},\ \Eprint
  {http://arxiv.org/abs/1212.2970} {arXiv:1212.2970 [hep-ph]} \BibitemShut
  {NoStop}%
\bibitem [{\citenamefont {Horns}\ \emph
  {et~al.}(2013{\natexlab{b}})\citenamefont {Horns}, \citenamefont {Lindner},
  \citenamefont {Lobanov},\ and\ \citenamefont {Ringwald}}]{Horns:2013ira}%
  \BibitemOpen
  \bibfield  {author} {\bibinfo {author} {\bibfnamefont {Dieter}\ \bibnamefont
  {Horns}}, \bibinfo {author} {\bibfnamefont {Axel}\ \bibnamefont {Lindner}},
  \bibinfo {author} {\bibfnamefont {Andrei}\ \bibnamefont {Lobanov}}, \ and\
  \bibinfo {author} {\bibfnamefont {Andreas}\ \bibnamefont {Ringwald}},\
  }\bibfield  {title} {\enquote {\bibinfo {title} {{WISPers from the Dark Side:
  Radio Probes of Axions and Hidden Photons}},}\ }in\ \href
  {http://inspirehep.net/record/1254430/files/arXiv:1309.4170.pdf} {\emph
  {\bibinfo {booktitle} {{9th Patras Workshop on Axions, WIMPs \& WISPs
  (PATRAS13) Mainz, Germany, June 24-28, 2013}}}}\ (\bibinfo {year} {2013})\
  \Eprint {http://arxiv.org/abs/1309.4170} {arXiv:1309.4170 [physics.ins-det]}
  \BibitemShut {NoStop}%
\bibitem [{\citenamefont {Stadnik}\ and\ \citenamefont
  {Flambaum}(2014)}]{Stadnik:2013raa}%
  \BibitemOpen
  \bibfield  {author} {\bibinfo {author} {\bibfnamefont {Y.~V.}\ \bibnamefont
  {Stadnik}}\ and\ \bibinfo {author} {\bibfnamefont {V.~V.}\ \bibnamefont
  {Flambaum}},\ }\bibfield  {title} {\enquote {\bibinfo {title} {{Axion-induced
  effects in atoms, molecules, and nuclei: Parity nonconservation, anapole
  moments, electric dipole moments, and spin-gravity and spin-axion momentum
  couplings}},}\ }\href {\doibase 10.1103/PhysRevD.89.043522} {\bibfield
  {journal} {\bibinfo  {journal} {Phys. Rev.}\ }\textbf {\bibinfo {volume}
  {D89}},\ \bibinfo {pages} {043522} (\bibinfo {year} {2014})},\ \Eprint
  {http://arxiv.org/abs/1312.6667} {arXiv:1312.6667 [hep-ph]} \BibitemShut
  {NoStop}%
\bibitem [{\citenamefont {Beck}(2013)}]{Beck:2013jha}%
  \BibitemOpen
  \bibfield  {author} {\bibinfo {author} {\bibfnamefont {Christian}\
  \bibnamefont {Beck}},\ }\bibfield  {title} {\enquote {\bibinfo {title}
  {{Possible resonance effect of axionic dark matter in Josephson
  junctions}},}\ }\href {\doibase 10.1103/PhysRevLett.111.231801} {\bibfield
  {journal} {\bibinfo  {journal} {Phys. Rev. Lett.}\ }\textbf {\bibinfo
  {volume} {111}},\ \bibinfo {pages} {231801} (\bibinfo {year} {2013})},\
  \Eprint {http://arxiv.org/abs/1309.3790} {arXiv:1309.3790 [hep-ph]}
  \BibitemShut {NoStop}%
\bibitem [{\citenamefont {Beck}(2015)}]{Beck:2014aqa}%
  \BibitemOpen
  \bibfield  {author} {\bibinfo {author} {\bibfnamefont {Christian}\
  \bibnamefont {Beck}},\ }\bibfield  {title} {\enquote {\bibinfo {title}
  {{Axion mass estimates from resonant Josephson junctions}},}\ }\href
  {\doibase 10.1016/j.dark.2015.03.002} {\bibfield  {journal} {\bibinfo
  {journal} {Phys. Dark Univ.}\ }\textbf {\bibinfo {volume} {7-8}},\ \bibinfo
  {pages} {6--11} (\bibinfo {year} {2015})},\ \Eprint
  {http://arxiv.org/abs/1403.5676} {arXiv:1403.5676 [hep-ph]} \BibitemShut
  {NoStop}%
\bibitem [{\citenamefont {Hong}\ \emph {et~al.}(2014)\citenamefont {Hong},
  \citenamefont {Kim}, \citenamefont {Nam},\ and\ \citenamefont
  {Semertzidis}}]{Hong:2014vua}%
  \BibitemOpen
  \bibfield  {author} {\bibinfo {author} {\bibfnamefont {Jooyoo}\ \bibnamefont
  {Hong}}, \bibinfo {author} {\bibfnamefont {Jihn~E.}\ \bibnamefont {Kim}},
  \bibinfo {author} {\bibfnamefont {Soonkeon}\ \bibnamefont {Nam}}, \ and\
  \bibinfo {author} {\bibfnamefont {Yannis}\ \bibnamefont {Semertzidis}},\
  }\bibfield  {title} {\enquote {\bibinfo {title} {{Calculations of resonance
  enhancement factor in axion-search tube-experiments}},}\ }\href@noop {} {\
  (\bibinfo {year} {2014})},\ \Eprint {http://arxiv.org/abs/1403.1576}
  {arXiv:1403.1576 [hep-ph]} \BibitemShut {NoStop}%
\bibitem [{\citenamefont {Roberts}\ \emph
  {et~al.}(2014{\natexlab{a}})\citenamefont {Roberts}, \citenamefont {Stadnik},
  \citenamefont {Dzuba}, \citenamefont {Flambaum}, \citenamefont {Leefer},\
  and\ \citenamefont {Budker}}]{Roberts:2014dda}%
  \BibitemOpen
  \bibfield  {author} {\bibinfo {author} {\bibfnamefont {B.~M.}\ \bibnamefont
  {Roberts}}, \bibinfo {author} {\bibfnamefont {Y.~V.}\ \bibnamefont
  {Stadnik}}, \bibinfo {author} {\bibfnamefont {V.~A.}\ \bibnamefont {Dzuba}},
  \bibinfo {author} {\bibfnamefont {V.~V.}\ \bibnamefont {Flambaum}}, \bibinfo
  {author} {\bibfnamefont {N.}~\bibnamefont {Leefer}}, \ and\ \bibinfo {author}
  {\bibfnamefont {D.}~\bibnamefont {Budker}},\ }\bibfield  {title} {\enquote
  {\bibinfo {title} {{Limiting P-odd interactions of cosmic fields with
  electrons, protons and neutrons}},}\ }\href {\doibase
  10.1103/PhysRevLett.113.081601} {\bibfield  {journal} {\bibinfo  {journal}
  {Phys. Rev. Lett.}\ }\textbf {\bibinfo {volume} {113}},\ \bibinfo {pages}
  {081601} (\bibinfo {year} {2014}{\natexlab{a}})},\ \Eprint
  {http://arxiv.org/abs/1404.2723} {arXiv:1404.2723 [hep-ph]} \BibitemShut
  {NoStop}%
\bibitem [{\citenamefont {Roberts}\ \emph
  {et~al.}(2014{\natexlab{b}})\citenamefont {Roberts}, \citenamefont {Stadnik},
  \citenamefont {Dzuba}, \citenamefont {Flambaum}, \citenamefont {Leefer},\
  and\ \citenamefont {Budker}}]{Roberts:2014cga}%
  \BibitemOpen
  \bibfield  {author} {\bibinfo {author} {\bibfnamefont {B.~M.}\ \bibnamefont
  {Roberts}}, \bibinfo {author} {\bibfnamefont {Y.~V.}\ \bibnamefont
  {Stadnik}}, \bibinfo {author} {\bibfnamefont {V.~A.}\ \bibnamefont {Dzuba}},
  \bibinfo {author} {\bibfnamefont {V.~V.}\ \bibnamefont {Flambaum}}, \bibinfo
  {author} {\bibfnamefont {N.}~\bibnamefont {Leefer}}, \ and\ \bibinfo {author}
  {\bibfnamefont {D.}~\bibnamefont {Budker}},\ }\bibfield  {title} {\enquote
  {\bibinfo {title} {{Parity-violating interactions of cosmic fields with
  atoms, molecules, and nuclei: Concepts and calculations for laboratory
  searches and extracting limits}},}\ }\href {\doibase
  10.1103/PhysRevD.90.096005} {\bibfield  {journal} {\bibinfo  {journal} {Phys.
  Rev.}\ }\textbf {\bibinfo {volume} {D90}},\ \bibinfo {pages} {096005}
  (\bibinfo {year} {2014}{\natexlab{b}})},\ \Eprint
  {http://arxiv.org/abs/1409.2564} {arXiv:1409.2564 [hep-ph]} \BibitemShut
  {NoStop}%
\bibitem [{\citenamefont {Stadnik}\ and\ \citenamefont
  {Flambaum}(2015)}]{Stadnik:2014tta}%
  \BibitemOpen
  \bibfield  {author} {\bibinfo {author} {\bibfnamefont {Y.~V.}\ \bibnamefont
  {Stadnik}}\ and\ \bibinfo {author} {\bibfnamefont {V.~V.}\ \bibnamefont
  {Flambaum}},\ }\bibfield  {title} {\enquote {\bibinfo {title} {{Searching for
  dark matter and variation of fundamental constants with laser and maser
  interferometry}},}\ }\href {\doibase 10.1103/PhysRevLett.114.161301}
  {\bibfield  {journal} {\bibinfo  {journal} {Phys. Rev. Lett.}\ }\textbf
  {\bibinfo {volume} {114}},\ \bibinfo {pages} {161301} (\bibinfo {year}
  {2015})},\ \Eprint {http://arxiv.org/abs/1412.7801} {arXiv:1412.7801
  [hep-ph]} \BibitemShut {NoStop}%
\bibitem [{\citenamefont {Hill}(2015{\natexlab{a}})}]{Hill:2015kva}%
  \BibitemOpen
  \bibfield  {author} {\bibinfo {author} {\bibfnamefont {Christopher~T.}\
  \bibnamefont {Hill}},\ }\bibfield  {title} {\enquote {\bibinfo {title}
  {{Axion Induced Oscillating Electric Dipole Moments}},}\ }\href {\doibase
  10.1103/PhysRevD.91.111702} {\bibfield  {journal} {\bibinfo  {journal} {Phys.
  Rev.}\ }\textbf {\bibinfo {volume} {D91}},\ \bibinfo {pages} {111702}
  (\bibinfo {year} {2015}{\natexlab{a}})},\ \Eprint
  {http://arxiv.org/abs/1504.01295} {arXiv:1504.01295 [hep-ph]} \BibitemShut
  {NoStop}%
\bibitem [{\citenamefont {Hill}(2015{\natexlab{b}})}]{Hill:2015vma}%
  \BibitemOpen
  \bibfield  {author} {\bibinfo {author} {\bibfnamefont {Christopher~T.}\
  \bibnamefont {Hill}},\ }\href@noop {} {\enquote {\bibinfo {title} {{Axion
  Induced Oscillating Electric Dipole Moment of the Electron}},}\ } (\bibinfo
  {year} {2015}{\natexlab{b}}),\ \Eprint {http://arxiv.org/abs/1508.04083}
  {arXiv:1508.04083 [hep-ph]} \BibitemShut {NoStop}%
\bibitem [{\citenamefont {McAllister}\ \emph {et~al.}(2015)\citenamefont
  {McAllister}, \citenamefont {Parker},\ and\ \citenamefont
  {Tobar}}]{McAllister:2015zcz}%
  \BibitemOpen
  \bibfield  {author} {\bibinfo {author} {\bibfnamefont {Ben~T.}\ \bibnamefont
  {McAllister}}, \bibinfo {author} {\bibfnamefont {Stephen~R.}\ \bibnamefont
  {Parker}}, \ and\ \bibinfo {author} {\bibfnamefont {Michael~E.}\ \bibnamefont
  {Tobar}},\ }\bibfield  {title} {\enquote {\bibinfo {title} {{Axion Dark
  Matter Coupling to Resonant Photons via Magnetic Field}},}\ }\href@noop {} {\
   (\bibinfo {year} {2015})},\ \Eprint {http://arxiv.org/abs/1512.05547}
  {arXiv:1512.05547 [hep-ph]} \BibitemShut {NoStop}%
\bibitem [{\citenamefont {Vogel}\ \emph {et~al.}(2013)\citenamefont {Vogel}
  \emph {et~al.}}]{Vogel:2013bta}%
  \BibitemOpen
  \bibfield  {author} {\bibinfo {author} {\bibfnamefont {J.~K.}\ \bibnamefont
  {Vogel}} \emph {et~al.},\ }\bibfield  {title} {\enquote {\bibinfo {title}
  {{IAXO - The International Axion Observatory}},}\ \ }(\bibinfo {year}
  {2013})\ \Eprint {http://arxiv.org/abs/1302.3273} {arXiv:1302.3273
  [physics.ins-det]} \BibitemShut {NoStop}%
\bibitem [{\citenamefont {Myers}\ \emph {et~al.}(2007)\citenamefont {Myers},
  \citenamefont {Slichter}, \citenamefont {Hatridge}, \citenamefont {Busch},
  \citenamefont {M{\"o}{\ss}le}, \citenamefont {McDermott}, \citenamefont
  {Trabesinger},\ and\ \citenamefont {Clarke}}]{myers2007calculated}%
  \BibitemOpen
  \bibfield  {author} {\bibinfo {author} {\bibfnamefont {Whittier}\
  \bibnamefont {Myers}}, \bibinfo {author} {\bibfnamefont {Daniel}\
  \bibnamefont {Slichter}}, \bibinfo {author} {\bibfnamefont {Michael}\
  \bibnamefont {Hatridge}}, \bibinfo {author} {\bibfnamefont {Sarah}\
  \bibnamefont {Busch}}, \bibinfo {author} {\bibfnamefont {Michael}\
  \bibnamefont {M{\"o}{\ss}le}}, \bibinfo {author} {\bibfnamefont {Robert}\
  \bibnamefont {McDermott}}, \bibinfo {author} {\bibfnamefont {Andreas}\
  \bibnamefont {Trabesinger}}, \ and\ \bibinfo {author} {\bibfnamefont {John}\
  \bibnamefont {Clarke}},\ }\bibfield  {title} {\enquote {\bibinfo {title}
  {Calculated signal-to-noise ratio of {MRI} detected with {SQUID}s and faraday
  detectors in fields from 10$\mu${T} to 1.5 {T}},}\ }\href@noop {} {\bibfield
  {journal} {\bibinfo  {journal} {Journal of Magnetic Resonance}\ }\textbf
  {\bibinfo {volume} {186}},\ \bibinfo {pages} {182--192} (\bibinfo {year}
  {2007})}\BibitemShut {NoStop}%
\bibitem [{\citenamefont {Clarke}\ \emph {et~al.}(1979)\citenamefont {Clarke},
  \citenamefont {Tesche},\ and\ \citenamefont {Giffard}}]{Clarke:1979aa}%
  \BibitemOpen
  \bibfield  {author} {\bibinfo {author} {\bibfnamefont {John}\ \bibnamefont
  {Clarke}}, \bibinfo {author} {\bibfnamefont {Claudia~D.}\ \bibnamefont
  {Tesche}}, \ and\ \bibinfo {author} {\bibfnamefont {R.P.}\ \bibnamefont
  {Giffard}},\ }\bibfield  {title} {\enquote {\bibinfo {title} {Optimization of
  dc squid voltmeter and magnetometer circuits},}\ }\href@noop {} {\bibfield
  {journal} {\bibinfo  {journal} {Journal of Low Temperature Physics}\ }\textbf
  {\bibinfo {volume} {37}},\ \bibinfo {pages} {405--420} (\bibinfo {year}
  {1979})}\BibitemShut {NoStop}%
\bibitem [{\citenamefont {Shoenberg}(1954)}]{shoenberg1954superconductivity}%
  \BibitemOpen
  \bibfield  {author} {\bibinfo {author} {\bibfnamefont {D}~\bibnamefont
  {Shoenberg}},\ }\href@noop {} {\emph {\bibinfo {title} {Superconductivity}}}\
  (\bibinfo  {publisher} {Cambridge University Press},\ \bibinfo {year}
  {1954})\BibitemShut {NoStop}%
\bibitem [{\citenamefont {Zimmerman}(1971)}]{zimmerman1971sensitivity}%
  \BibitemOpen
  \bibfield  {author} {\bibinfo {author} {\bibfnamefont {JE}~\bibnamefont
  {Zimmerman}},\ }\bibfield  {title} {\enquote {\bibinfo {title} {Sensitivity
  enhancement of superconducting quantum interference devices through the use
  of fractional-turn loops},}\ }\href@noop {} {\bibfield  {journal} {\bibinfo
  {journal} {Journal of Applied Physics}\ }\textbf {\bibinfo {volume} {42}},\
  \bibinfo {pages} {4483--4487} (\bibinfo {year} {1971})}\BibitemShut {NoStop}%
\bibitem [{\citenamefont {Drung}\ \emph {et~al.}(1990)\citenamefont {Drung},
  \citenamefont {Cantor}, \citenamefont {Peters}, \citenamefont {Scheer},\ and\
  \citenamefont {Koch}}]{drung1990low}%
  \BibitemOpen
  \bibfield  {author} {\bibinfo {author} {\bibfnamefont {D}~\bibnamefont
  {Drung}}, \bibinfo {author} {\bibfnamefont {R}~\bibnamefont {Cantor}},
  \bibinfo {author} {\bibfnamefont {M}~\bibnamefont {Peters}}, \bibinfo
  {author} {\bibfnamefont {HJ}~\bibnamefont {Scheer}}, \ and\ \bibinfo {author}
  {\bibfnamefont {H}~\bibnamefont {Koch}},\ }\bibfield  {title} {\enquote
  {\bibinfo {title} {Low-noise high-speed dc superconducting quantum
  interference device magnetometer with simplified feedback electronics},}\
  }\href@noop {} {\bibfield  {journal} {\bibinfo  {journal} {Applied Physics
  Letters}\ }\textbf {\bibinfo {volume} {57}},\ \bibinfo {pages} {406--408}
  (\bibinfo {year} {1990})}\BibitemShut {NoStop}%
\bibitem [{\citenamefont {Anton}\ \emph {et~al.}(2013)\citenamefont {Anton},
  \citenamefont {Birenbaum}, \citenamefont {O'Kelley}, \citenamefont
  {Bolkhovsky}, \citenamefont {Braje}, \citenamefont {Fitch}, \citenamefont
  {Neeley}, \citenamefont {Hilton}, \citenamefont {Cho}, \citenamefont {Irwin}
  \emph {et~al.}}]{anton2013magnetic}%
  \BibitemOpen
  \bibfield  {author} {\bibinfo {author} {\bibfnamefont {SM}~\bibnamefont
  {Anton}}, \bibinfo {author} {\bibfnamefont {JS}~\bibnamefont {Birenbaum}},
  \bibinfo {author} {\bibfnamefont {SR}~\bibnamefont {O'Kelley}}, \bibinfo
  {author} {\bibfnamefont {V}~\bibnamefont {Bolkhovsky}}, \bibinfo {author}
  {\bibfnamefont {DA}~\bibnamefont {Braje}}, \bibinfo {author} {\bibfnamefont
  {G}~\bibnamefont {Fitch}}, \bibinfo {author} {\bibfnamefont {M}~\bibnamefont
  {Neeley}}, \bibinfo {author} {\bibfnamefont {GC}~\bibnamefont {Hilton}},
  \bibinfo {author} {\bibfnamefont {H-M}\ \bibnamefont {Cho}}, \bibinfo
  {author} {\bibfnamefont {KD}~\bibnamefont {Irwin}},  \emph {et~al.},\
  }\bibfield  {title} {\enquote {\bibinfo {title} {Magnetic flux noise in {DC}
  {SQUID}s: Temperature and geometry dependence},}\ }\href@noop {} {\bibfield
  {journal} {\bibinfo  {journal} {Physical Review Letters}\ }\textbf {\bibinfo
  {volume} {110}},\ \bibinfo {pages} {147002} (\bibinfo {year}
  {2013})}\BibitemShut {NoStop}%
\bibitem [{\citenamefont {Seton}\ \emph {et~al.}(1995)\citenamefont {Seton},
  \citenamefont {Bussell}, \citenamefont {Hutchison},\ and\ \citenamefont
  {Lurie}}]{seton1995use}%
  \BibitemOpen
  \bibfield  {author} {\bibinfo {author} {\bibfnamefont {HC}~\bibnamefont
  {Seton}}, \bibinfo {author} {\bibfnamefont {DM}~\bibnamefont {Bussell}},
  \bibinfo {author} {\bibfnamefont {JMS}\ \bibnamefont {Hutchison}}, \ and\
  \bibinfo {author} {\bibfnamefont {DJ}~\bibnamefont {Lurie}},\ }\bibfield
  {title} {\enquote {\bibinfo {title} {{Use of a DC SQUID receiver preamplifier
  in a low field MRI system}},}\ }\href@noop {} {\bibfield  {journal} {\bibinfo
   {journal} {Applied Superconductivity, IEEE Transactions on}\ }\textbf
  {\bibinfo {volume} {5}},\ \bibinfo {pages} {3218--3221} (\bibinfo {year}
  {1995})}\BibitemShut {NoStop}%
\bibitem [{\citenamefont {Seton}\ \emph {et~al.}(1999)\citenamefont {Seton},
  \citenamefont {Hutchison},\ and\ \citenamefont
  {Bussell}}]{seton1999gradiometer}%
  \BibitemOpen
  \bibfield  {author} {\bibinfo {author} {\bibfnamefont {HC}~\bibnamefont
  {Seton}}, \bibinfo {author} {\bibfnamefont {JMS}\ \bibnamefont {Hutchison}},
  \ and\ \bibinfo {author} {\bibfnamefont {DM}~\bibnamefont {Bussell}},\
  }\bibfield  {title} {\enquote {\bibinfo {title} {{Gradiometer pick-up coil
  design for a low field SQUID-MRI system}},}\ }\href@noop {} {\bibfield
  {journal} {\bibinfo  {journal} {Magnetic Resonance Materials in Physics,
  Biology and Medicine}\ }\textbf {\bibinfo {volume} {8}},\ \bibinfo {pages}
  {116--120} (\bibinfo {year} {1999})}\BibitemShut {NoStop}%
\bibitem [{\citenamefont {Bonaldi}\ \emph {et~al.}(1999)\citenamefont
  {Bonaldi}, \citenamefont {Falferi}, \citenamefont {Cerdonio}, \citenamefont
  {Vinante}, \citenamefont {Dolesi},\ and\ \citenamefont
  {Vitale}}]{bonaldi1999thermal}%
  \BibitemOpen
  \bibfield  {author} {\bibinfo {author} {\bibfnamefont {Michele}\ \bibnamefont
  {Bonaldi}}, \bibinfo {author} {\bibfnamefont {Paolo}\ \bibnamefont
  {Falferi}}, \bibinfo {author} {\bibfnamefont {Massimo}\ \bibnamefont
  {Cerdonio}}, \bibinfo {author} {\bibfnamefont {Andrea}\ \bibnamefont
  {Vinante}}, \bibinfo {author} {\bibfnamefont {Rita}\ \bibnamefont {Dolesi}},
  \ and\ \bibinfo {author} {\bibfnamefont {Stefano}\ \bibnamefont {Vitale}},\
  }\bibfield  {title} {\enquote {\bibinfo {title} {Thermal noise in a high {Q}
  cryogenic resonator},}\ }\href@noop {} {\bibfield  {journal} {\bibinfo
  {journal} {Review of scientific instruments}\ }\textbf {\bibinfo {volume}
  {70}},\ \bibinfo {pages} {1851--1856} (\bibinfo {year} {1999})}\BibitemShut
  {NoStop}%
\bibitem [{\citenamefont {Matlachov}\ \emph {et~al.}(2004)\citenamefont
  {Matlachov}, \citenamefont {Volegov}, \citenamefont {Espy}, \citenamefont
  {George},\ and\ \citenamefont {Kraus}}]{matlachov2004squid}%
  \BibitemOpen
  \bibfield  {author} {\bibinfo {author} {\bibfnamefont {Andrei~N}\
  \bibnamefont {Matlachov}}, \bibinfo {author} {\bibfnamefont {Petr~L}\
  \bibnamefont {Volegov}}, \bibinfo {author} {\bibfnamefont {Michelle~A}\
  \bibnamefont {Espy}}, \bibinfo {author} {\bibfnamefont {John~S}\ \bibnamefont
  {George}}, \ and\ \bibinfo {author} {\bibfnamefont {Robert~H}\ \bibnamefont
  {Kraus}},\ }\bibfield  {title} {\enquote {\bibinfo {title} {{SQUID} detected
  {NMR} in microtesla magnetic fields},}\ }\href@noop {} {\bibfield  {journal}
  {\bibinfo  {journal} {Journal of Magnetic Resonance}\ }\textbf {\bibinfo
  {volume} {170}},\ \bibinfo {pages} {1--7} (\bibinfo {year}
  {2004})}\BibitemShut {NoStop}%
\bibitem [{\citenamefont {McDermott}\ \emph {et~al.}(2004)\citenamefont
  {McDermott}, \citenamefont {Lee}, \citenamefont {Ten~Haken}, \citenamefont
  {Trabesinger}, \citenamefont {Pines},\ and\ \citenamefont
  {Clarke}}]{mcdermott2004microtesla}%
  \BibitemOpen
  \bibfield  {author} {\bibinfo {author} {\bibfnamefont {Robert}\ \bibnamefont
  {McDermott}}, \bibinfo {author} {\bibfnamefont {SeungKyun}\ \bibnamefont
  {Lee}}, \bibinfo {author} {\bibfnamefont {Bennie}\ \bibnamefont {Ten~Haken}},
  \bibinfo {author} {\bibfnamefont {Andreas~H}\ \bibnamefont {Trabesinger}},
  \bibinfo {author} {\bibfnamefont {Alexander}\ \bibnamefont {Pines}}, \ and\
  \bibinfo {author} {\bibfnamefont {John}\ \bibnamefont {Clarke}},\ }\bibfield
  {title} {\enquote {\bibinfo {title} {Microtesla {MRI} with a superconducting
  quantum interference device},}\ }\href@noop {} {\bibfield  {journal}
  {\bibinfo  {journal} {Proceedings of the National Academy of Sciences of the
  United States of America}\ }\textbf {\bibinfo {volume} {101}},\ \bibinfo
  {pages} {7857--7861} (\bibinfo {year} {2004})}\BibitemShut {NoStop}%
\bibitem [{\citenamefont {Lee}\ and\ \citenamefont
  {Romalis}(2008)}]{lee2008calculation}%
  \BibitemOpen
  \bibfield  {author} {\bibinfo {author} {\bibfnamefont {S-K}\ \bibnamefont
  {Lee}}\ and\ \bibinfo {author} {\bibfnamefont {MV}~\bibnamefont {Romalis}},\
  }\bibfield  {title} {\enquote {\bibinfo {title} {Calculation of magnetic
  field noise from high-permeability magnetic shields and conducting objects
  with simple geometry},}\ }\href@noop {} {\bibfield  {journal} {\bibinfo
  {journal} {Journal of Applied Physics}\ }\textbf {\bibinfo {volume} {103}},\
  \bibinfo {pages} {084904} (\bibinfo {year} {2008})}\BibitemShut {NoStop}%
\bibitem [{Rom()}]{RomalisPrivate}%
  \BibitemOpen
  \href@noop {} {}\bibinfo {howpublished} {{M}ike {R}omalis, private
  communication.}\BibitemShut {Stop}%
\end{thebibliography}%

\end{document}